\documentclass[journal,10pt]{IEEEtran}

\usepackage{booktabs}

\usepackage{color}
\usepackage{graphicx}
\graphicspath{{figs/}}
\usepackage{subfigure}
\usepackage{amssymb}
\usepackage{amsmath}
\DeclareMathOperator*{\argmax}{arg\,max} 
\usepackage{algorithm}
\usepackage{algorithmic}

\usepackage{array}
\usepackage{amsmath}
\usepackage{amsmath,graphicx,epsfig,float,times,latexsym,verbatim,mathrsfs,amssymb,textcomp,txfonts,color,setspace,cite}
\usepackage{algorithm}
\usepackage{algorithmic}
\usepackage{url}

\begin{document}
\title{Fast Beam Tracking for Reconfigurable Intelligent Surface Assisted Mobile mmWave Networks}

\author{Xiaowen Tian and 
		Zhi Sun
\thanks{This work was supported by the
National Science Foundation (Grant No. 1652502).}
\thanks{Xiaowen Tian and Zhi Sun are with the Department of Electrical Engineering, University at Buffalo, Buffalo, NY 14260 USA (E-mail: xiaowent@buffalo.edu; zhisun@buffalo.edu).}
\thanks{This work has been submitted to the Elsevier Computer Networks for possible publication. Copyright may be transferred without notice, after which this version may no longer be accessible.}
}

\maketitle
\thispagestyle{plain}

\begin{abstract}
Millimeter wave (mmWave) communications are vulnerable to blockages and node mobility due to the highly directional signal beams.
The emerging Reconfigurable Intelligent Surfaces (RISs) technique can effectively mitigate the blockage problem by exploring the non-line-of-sight (NLOS) path, where the beam switching is realized by digitally configuring the phases of RIS elements.
To date, most efforts have been made in the stationary scenario.
However, when considering node mobility, beam tracking algorithms designed specifically for RIS are needed in order to maintain the NLOS link.
In this paper, a fast RIS-based beam tracking algorithm is developed by partly transforming the large amount of signaling time into the calculation happens at base station in a mmWave system with mobile users.
Specifically, the differential form of optimal RIS configuration is exploited as the updating beam tracking parameter to avoid complex channel estimation procedure.
The RIS-based beam tracking problem is then transformed into an optimization problem whose solution is found by a calculation-based search.
Finally, by training on a small set candidate, RIS-based beam tracking is realized. 
The effectiveness and efficiency of the proposed RIS-based beam tracking algorithm is evaluated by simulations.
It shows that the proposed algorithm has near-optimal performance with dramatic savings in terms of signaling time.
\end{abstract}


\pagestyle{plain}

\vspace{-0.0 cm}
\section{Introduction}
\label{sec: intro}
\pagestyle{plain}


Operating at the vast unused millimeter wave spectrum, millimeter wave (mmWave) systems have become a promising technique in the fifth-generation (5G) communications to overcome the spectrum congestion problem.
Combining with beamforming techniques in multiple antenna systems, mmWave signals are of high directionality, i.e., having much narrower beams, because of its much shorter wavelength comparing with conventional below 6GHz signals \cite{mmWave1}.
Therefore, despite the advantages, mmWave signals are sensitive to blockage and node mobility at the same time \cite{mmWave2}.

The Reconfigurable Intelligent Surfaces (RISs) \cite{RIS1}-\cite{RIS3}, working by digitally configuring the phases of the large number of elements on them, offer a solution to the problem of blockage in mmWave systems.
It is usually believed that the wireless communication channels are fixed and can only be compensated by the design techniques and signal processing procedures at the transceiver sides.
However, RISs rise as the key of the roadmap to smart radio environment because that they can be programmed to reflect the electromagnetic waves to the unnatural directions \cite{Smart Radio}.
By adding an RIS to the point-to-point mmWave communication system, an additional non-light-of-sigh (NLOS) path/channel is offered as is shown in our system model in Fig. \ref{fig: system model}.
The effect of manipulating the wireless communication channels by configuring the phases of RISs is proved by experiments on prototypes \cite{RISxin}-\cite{RISDOCOMO}, which provides the foundation of the multiple literatures on designing RISs in mmWave systems in order to overcome the challenge of blockage in mmWave systems.

Although adding a RIS into the mmWave systems creates an additional path and overcome the blockage issue, the support for mobile users is also essential, especially with the rapid developments of autonomous vehicle techniques \cite{btguo}, \cite{btwang}.
One solution to user mobility in mmWave systems is adopting beam tracking algorithms \cite{btinfo2017}-\cite{btrl}, which automatically switch the paired beams in order to maintain the quality of communication link above a certain threshold.
However, existing beam tracking algorithms in mmWave systems can not be used directly with RIS-assisted systems.
For example, turning on one element of RIS \cite{btinfo2017} is not able to have large enough received signal strength to be detected.
In addition, the large number of RIS elements will deteriorate the performance of Kalman filter-based tracking algorithms \cite{GlobalSIP}, or make the complexity of beam tracking algorithms intolerable \cite{11ad}.
Therefore, beam tracking algorithms designed specifically for RIS-assisted mmWave systems are needed.

One natural idea for RIS-based beam tracking algorithms is to estimate the AP-RIS-UE channel first and configure the RIS elements accordingly.
However, many literature on RIS-assisted channel estimation considers the static case, where the positions of the three parties remain unchanged \cite{RIS12}-\cite{RIS14}.
When considering mobility case, where the AP-RIS-UE channel is time-changing, those estimation procedures are more frequently called, increasing the complexity of beam tracking algorithms dramatically. 
Therefore, channel estimation should be avoided when designing RIS-based beam tracking algorithms.
In \cite{RIS11}, channel parameters are updated using extended Kalman filter, which is also utilized in systems without RISs \cite{GlobalSIP}.
Drawback of such directly utilization without adaptation is that large antenna array will deteriorate the accuracy of beam tracking because of beam misalignment.
Novel layout of the RIS-assisted system is proposed in \cite{RIS15}, where the RIS is attached on the vehicle.
In \cite{RIS15}, the unchanging Doppler frequency is exploited to estimate the time-changing cascaded channel.
Unfortunately, it is not in mmWave system and it exploits the AP-UE scatter-rich channel, at the same time still involves multiple training time-slots.

In conclusion, RIS-based beam tracking algorithms when considering node mobility in mmWave networks need to be specifically designed, whose challenges are listed as follows.
\begin{itemize}

\item RIS is passively reflecting signals.
The fact that RISs have no signal processing ability makes the communication involve three parties.
In other words, the configuration of the large amount of RIS elements can only be done based on the received signals, which complicates the RIS-based beam tracking algorithms since the algorithms without considering RIS can not be applied directly.

\item Finding best beam pairing solution is cumbersome.
MmWave signals have narrower beams.
Conventionally, e.g., in 11ad protocol, the beam alignment is done by switching the antenna to omni-directional mode at one side and remaining narrow beam mode at the other.
By exhaustively changing the narrow beam directions at both sides, a best spatial narrow beam pair which has the best communication performance is found.
In RIS-assisted systems, such complexity raises from square of the number of spatial beams into the cube of it, which is usually unacceptable.

\item RIS-based beam tracking algorithms need to be effective and efficient.
Channel parameters other than the complete channel state information (CSI) are to be updated in efficient beam switching algorithms since such procedure is called more frequently in order to maintain the communication link quality.
On the contrary, conventional channel estimation procedure is called only after severe signal loss.
Therefore, accuracy of updating the channel parameters without having to estimate the complete CSI is also a big challenge.

\end{itemize}
In this paper, we design the beam tracking algorithm particularly for RIS-assisted mmWave systems considering node mobility.
We propose to avoid the complex channel estimation procedure by exploiting the channel difference between adjacent user positions to obtain the update information to configure the RIS.
We also manage to partly transform the time-consuming signaling time into calculation complexity happens at the base station.
The major contributions of this paper are summarized as follows.
\begin{itemize}

\item We propose the system which utilizes RIS to create an additional NLOS path in order to overcome blockage issue and support node mobility at the same time in mmWave systems.
We formulate the beam tracking problem and have the expression of optimal RIS configuration, which will be exploited in differential form, serving as the parameters for updating during beam tracking procedure.

\item We propose an effective beam tracking algorithm which transforms the high time complexity of beam searching procedure into calculation complexity at base station side.
We transform the RIS beam configuration problem into an optimization problem by exploiting both the received signal strength (RSS) ratio and received signal angle difference of the optimal RIS configuration status and the current status.
By solving the transformed optimization problem using a two-dimensional grid-based search, a small-size candidate set of RIS configurations is found.
Then by applying this small set downlink training, signaling time is saved dramatically.

\item Simulation studies prove the effectiveness as well as the low complexity of our proposed RIS-based beam tracking algorithm in mmWave systems by comparing with conventional exhaustive searching strategy.
It is shown that our proposed algorithm has near-optimal performance with dramatic savings in terms of signaling time.

\end{itemize}

The remainder of this paper is organized as follows.
The related work is discussed in Sec. \ref{sec: related work}.
Sec. \ref{sec: two} describes the system model when applying RIS into mmWave systems and formulates the beam tracking problem.
The optimal RIS configuration with respect to channel information is also provided.
In Sec. \ref{sec: mobility}, we propose our RIS-based beam tracking algorithm by obtaining the update rule and proposing the channel complex gain model.
Then we transform the RIS-based beam tracking problem into an optimization problem by exploiting the received signal strength and angles comparing to optimal configuration case.
A search-based calculation returns a small set candidates and the overall signaling procedure is proposed then.
In Sec. \ref{sec: sim}, we explain how the channel is generated when considering a mobile user followed by multiple simulation studies conducted to verify our proposed algorithm.
Finally, conclusions are drawn in Sec. \ref{sec: conclusion}.

Notations:
Boldface lower-case and upper-case letters indicate column vectors and matrices, respectively. 
$(\cdot)^H$ denotes the transpose-conjugate operation.
$\mathbb{E} \{ \cdot \}$ represents statistical expectation. 
$\angle \{ \cdot \}$ represents the phase of a complex number.
$\mathbb{C}$ denotes the set of complex numbers.
$| \mathbf{A} |$ denotes the determinant of matrix $\mathbf{A}$.
$| a |$ and $| \mathbf{a} |$ are the magnitude and norm of a scalar $a$ and vector $\mathbf{a}$, respectively.  

\section{Related Work}
\label{sec: related work}

\subsection{Beam Tracking Algorithms in mmWave Systems}
Beam tracking is needed in mmWave systems because of its narrow beams and the algorithms can be divided by two catagories. 
One is to estimate the channel, the other is to predict the channel.
However, most of those algorithms can not be directly applied to RIS-assisted systems due to its unique characteristics as is mentioned in Sec. \ref{sec: intro}.
For example, in \cite{btinfo2017}, the authors assume omni-directional transmission which is unpractical to be directly applied to RIS because it's impossible to observe the low-power received signal with only one element of RIS working.
Similar to 802.11ad protocol \cite{11ad}, exhaustive searching strategy can be utilized in RIS-assisted systems but will induce intolerable complexity.
When designing RIS-based beam tracking algorithms, decreasing beam alignment complexity by narrowing down the search range is doable.
Same idea can be found in \cite{MobiWac} without considering RIS configuration, where the authors exploit the beam patterns of adjacent positions of user to decrease the search time.
Kalman filter is useful when designing beam tracking algorithms as in \cite{GlobalSIP} without considering RIS.
However, without adapting it in RIS-assisted networks, the accuracy of the algorithm decreases with larger number of RIS elements.
Learning-based beam tracking algorithms are emerging these days thanks to their ability in beam prediction \cite{btguo}-\cite{btrl}.
The machine-learning-based algorithms will require additional information, such as multiple base stations in \cite{btguo} and situational awareness information in \cite{btwang}, to train the learning model and execute beam prediction.
In addition, the reinforcement-learning-based algorithms, such as in \cite{btrl}, also require many times of failures before successfully selecting the next state action.

\subsection{RIS-based Channel Estimation Algorithms}

Since beam tracking is highly related to channel estimation and the participation of RIS will introduce unique challenges, RIS-based channel estimation algorithms are of importance in inspiring RIS-based beam tracking algorithms.
Since the passive behaviour and large number of RIS elements make RIS-based channel estimation more complex than in conventional systems, many literature focus on saving the signaling time by exploiting the unique channel properties.
In \cite{RIS12}, compressed sensing and deep learning based algorithm are proposed in order to obtain the channel state information with RIS systems.
In \cite{RIS13}, the channel feedback is exploited in order to extract the essential information needed to configure the RIS.
The complexity is transformed from searching exhaustively into selecting a codeword from a pre-set codebook that representing the unknown channel.
In \cite{RIS14}, the authors propose a manifold optimization (MO)-based algorithm by exploiting the inherent structure of the effective mmWave channel to optimally configure the RIS.
In order to decrease the complexity of obtaining complete CSI, algorithms which update channel parameters are commonly used when considering beam tracking problem.
In \cite{RIS11}, the RIS-based beam tracking algorithm is based on hierarchically searching for the auxiliary beams as the estimated cascaded channel.
With estimated channel parameters, the extended Kalman filter is utilized to update the estimated channel parameters and the RIS configuration.
However, the tracking algorithm will become inaccurate when the number of RIS elements is large because of the narrower beam misalignment \cite{GlobalSIP}.
With time-changing channels when considering node mobility, there are some parameters that are not changing, which can be exploited to estimate the channel.
In \cite{RIS15}, the RIS is implemented in the high-speed vehicle to generate slow fading RIS-UE channel.
The authors exploit the Doppler frequency to estimate the cascaded AP-RIS-UE channel followed by designing the RIS configuration in order to constructively align with the direct AP-UE channel.

In summary, as is mentioned in the challenges of our topic in Sec. \ref{sec: intro}, despite the mature studies of beam tracking algorithms in mmWave systems, they can not be directly applied to RIS systems because of its passive-reflecting characteristic and large number of elements.
In addition, most of the existing work about RIS configurations focus on the static scenarios and will require the procedure of channel estimation, which is of high complexity if used in beam tracking scenarios.
Therefore, in this paper, we propose the beam tracking algorithm designed specifically for RIS-assisted mmWave systems in order to overcome the issues of blockages and node mobility.  
By exploiting the signal strengths and angles of adjacent status received signals, our proposed solution can bypass the complex channel estimation procedure and update the beams based on the differential form of optimal RIS configuration, thus saving large amount of signaling time as well as maintaining the communication quality above a given threshold.

\section{System Model, Problem Formulation and Optimal Initial Configuration}
\label{sec: two}

In this section, we first describe the mmWave system model with a third-party RIS and formulate the RIS beam tracking problem under node mobility.
Then the optimal initial RIS configuration obtained by a searching manner is introduced based on our derivation of optimal RIS configuration represented by channel angular information.

\subsection{System Model}
\label{sec: model}

\begin{figure}[!t]
\centering
\vspace{-0.0 cm}
\includegraphics[width= 3 in]{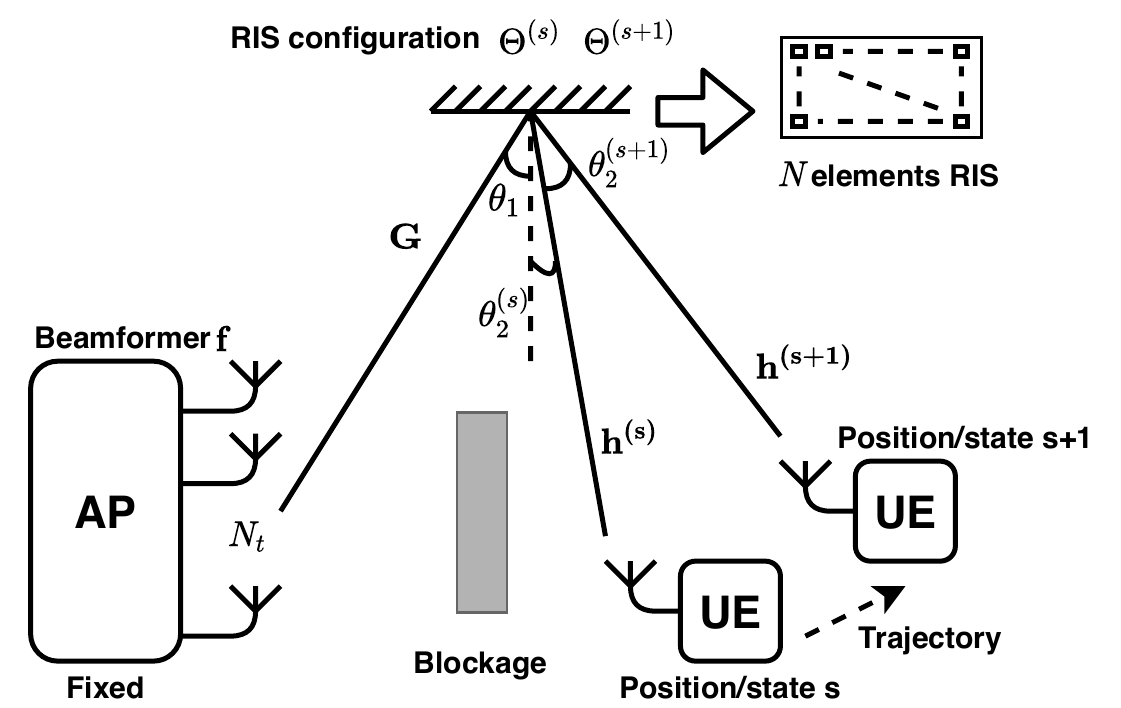}\vspace{-0.0 cm}
\caption{A mmWave system with a multiple antennas AP, a single antenna mobile UE and an $N$-element third party RIS. The LOS path is blocked and an additional NLOS path is created with the application of RIS.}
\label{fig: system model}
\vspace{-0.3 cm}
\end{figure}

Consider a mmWave system with a RIS as illustrated in Fig. \ref{fig: system model}, in which the AP is equipped with $N_t$ antennas and transmits the information to a single-antenna UE through beamformer $\mathbf{f} \in \mathbb{R}^{N_t \times 1}$.
Here we consider the case where the line-of-sight (LOS) path is blocked and AP and UE are communicating through the RIS, which has $N$ phase shifters to manipulate the signal's phases.
The transmitted symbols first goes through the AP-RIS channel $\mathbf{G} \in \mathbb{R}^{N \times N_t}$, and reflected by the RIS, whose effect is denoted by a diagonal matrix $\mathbf{\Theta} = \mathbf{diag}\{e^{j \phi_1}, e^{j \phi_2}, ..., e^{j \phi_N}\} \in \mathbb{R}^{N \times N}$.
Then the reflected signal goes through the RIS-UE channel $\mathbf{h}^{(s)} \in \mathbb{R}^{N \times 1}$, where the notation $s$ represents the $s$-th position/state, and finally received by UE.
Therefore, the received signal at position/state $s$ has the following form:
\begin{equation}
y_{DL}^{(s)} = (\mathbf{h}^{(s)})^H \mathbf{\Theta}^{(s)} \mathbf{G}^{(s)} \mathbf{f} x^{(s)} + n^{(s)}.
\label{eq: yDL}
\end{equation}
where $\mathbb{E}\{x^{(s)}(x^{(s)})^H\} = 1$ is the transmitted symbol and $n^{(s)} \sim \mathcal{CN}(0, \sigma_n^2)$ is additive white Gaussian noise.
Specifically, 
\begin{equation}
\mathbf{G} = \alpha \mathbf{a}_{RA}(\theta_1) \mathbf{a}_{AP}^H(\phi_{AP}),
\end{equation}
\begin{equation}
(\mathbf{h}^{(s)})^H = \beta_s \mathbf{a}_{RA}^H(\theta_2^{(s)}).
\end{equation}
where $\alpha$ and $\beta_s$ are channel coefficients.
$\mathbf{a}_{AP}(\phi_{AP})$, $\mathbf{a}_{RA}(\theta_1)$ and $\mathbf{a}_{RA}(\theta_2^{(s)})$ are steering vectors of the corresponding channels, with $\phi_{AP}$, $\theta_1$ and $\theta_2^{(s)}$ as AoD of AP-RIS channel, AoA of AP-RIS channel, and AoD of RIS-UE channel.
The steering vectors has the form of:
\begin{equation}
\mathbf{a}(\theta) = [1, e^{-j\frac{2 \pi d}{\lambda} \sin\theta}, ..., e^{-j\frac{2 \pi d}{\lambda} (N-1) \sin\theta}]^T.
\label{eq: steering vector}
\end{equation}

\subsection{Problem Formulation}
\label{sec: formulation}

According to the received signal in (\ref{eq: yDL}), the problem of optimal RIS configuration has the following form:
\begin{equation}
\mathbf{\Theta}^* = \argmax ~~| (\mathbf{h}^{(s)})^H \mathbf{\Theta} \mathbf{G} \mathbf{f}|^2.
\label{eq: obj ori}
\end{equation}
Our goal is to propose an RIS beam tracking algorithm in order to achieve high data rate while keeping the overhead as low as possible.
In order to design the RIS configuration during node mobility, an optimal initial configuration is introduced in the next subsection.

\subsection{Optimal Initial Configuration}
\label{sec: optimal conf}

Initial access is an essential step in mmWave communications in order to align the narrow beams of the transceivers.
In a mmWave system with RIS, the AP-RIS channel can be viewed as unchanged since the positions of both AP and RIS are fixed.
Therefore, in our paper, we focus on the RIS configuration under the assumption of $\mathbf{f} = \sqrt{\mathrm{SNR}} \times \frac{\mathbf{a}_{AP}(\phi_{AP})}{|\mathbf{a}_{AP}(\phi_{AP})|}$ in the following deduction.
Such assumption is reasonable because the AoD can be obtained either by the a 802.11ad beam searching procedure or by other AoD estimation algorithms.


With that being said, we can rewrite the received signal at UE from (\ref{eq: yDL}) as:
\begin{eqnarray}
y_{DL}^{(s)} \hspace{-0.3 cm}&=&\hspace{-0.3 cm} (\beta_s \mathbf{a}_{RA}^H(\theta_2^{(s)}) ) \mathbf{\Theta}^{(s)} (\alpha \mathbf{a}_{RA}(\theta_1) \mathbf{a}_{AP}^H(\phi_{AP})) \nonumber\\
\hspace{-0.3 cm}&~~~~&\hspace{-0.3 cm} \times  \sqrt{\mathrm{SNR}} \times \frac{\mathbf{a}_{AP}(\phi_{AP})}{|\mathbf{a}_{AP}(\phi_{AP})|} x^{(s)} + n^{(s)} \nonumber\\
\hspace{-0.3 cm}&~~~~&\hspace{-0.3 cm} = c \alpha \beta_s \mathbf{a}_{RA}^H(\theta_2^{(s)}) \mathbf{\Theta}^{(s)} \mathbf{a}_{RA}(\theta_1) x^{(s)} + n^{(s)},
\end{eqnarray}
where $c \triangleq \sqrt{SNR} |\mathbf{a}_{AP}(\phi_{AP})|$.
Ignoring noise, we can rewrite the objective function in (\ref{eq: obj ori}) as:
\begin{equation}
\{ \phi_1, ..., \phi_N \} = \argmax ~ |\beta_s \mathbf{a}^H_{RA}(\theta_2^{(s)}) \mathbf{\Theta}^{(s)} \mathbf{a}_{RA}(\theta_1)|^2.
\label{eq: obj ignore turbulance}
\end{equation}
Replacing the steering vectors with the form of (\ref{eq: steering vector}), the objective function (\ref{eq: obj ignore turbulance}) can be rewritten as:
\begin{equation}
\{ \phi_1^*, ..., \phi_N^* \} = \argmax ~ |\beta_s|^2 |\sum_{n=1}^{N} e^{j \left(\phi_n - \frac{2 \pi d}{\lambda} (n-1) (\sin\theta_1-\sin\theta_2^{(s)})\right) }|^2.
\end{equation}
In order to have the maximum value of the summation form, it is obvious to align the phases of RIS elements with the channel.
Therefore, we can have the optimal RIS configuration as:
\begin{equation}
(\phi_n^{(s)})^* = \begin{cases}
                0, \mathrm{if} ~ n = 1, \\
                \frac{2 \pi d}{\lambda} (\sin\theta_1-\sin\theta_2^{(s)}), \mathrm{if} ~ n = 2, \\
                (n-1) \phi_2^{(s)}, \mathrm{if} ~ n = 3, ..., N.
\end{cases}
\label{eq: optimal conf}
\end{equation}

Taking a closer look at the optimal RIS configuration (\ref{eq: optimal conf}), the term $(\sin\theta_1-\sin\theta_2^{(s)})$ is unknown to us since we can not extract the exact AoD and AoA information of the channel related to RIS.
Another observation is that as long as we set the phase of one RIS element, i.e, $\phi_2$, we can have the optimal phases for other elements.
Inspired by the above two observations, we propose the optimal initial RIS configuration $\mathbf{\Theta}^{(1)}$ in an exhaustive searching manner, which searches the value of $\phi_2 \in (0^\circ, 360^\circ)$ with a given searching step ($1^\circ$ in simulation studies).

\section{Fast RIS-Based Beam Tracking Algorithm}
\label{sec: mobility}

After obtaining the optimal initial RIS configuration $\mathbf{\Theta}^{(1)}$ and starting the downlink data transmission with fixed AP-RIS beamformer $\mathbf{f}$, RIS beam tracking algorithm is executed in order to maintain the link under node mobility.
In this section, we first derive the RIS configuration update rule consisting the differential form of a channel parameter.
Then, we propose a channel complex gain model in order to decouple the effect of RIS misalignment from the observation of a  received signal strength drop.
Next, we transform the beam tracking problem into an optimization problem by exploiting the received signal strength and angle.
Using a two-dimensional grid-based search, we find a small-size candidate set of RIS configurations.
Finally, we explain the whole procedure including the signaling procedure and the framework of the RIS-based fast beam tracking algorithm.

\subsection{State Transformation and Update Rule}
\label{sec: state}

We first define a \textit{state} during downlink data transmission as follows:
If during time-slots $t_n$, there is $|y_{DL}^{(t_n)}|^2\geq \gamma$, where $\gamma$ is a communication quality threshold, we say these time-slots are in the same state.
In the same state/status, the configuration of RIS for downlink data transmission remains the same.
This is reasonable since we don't have to update the RIS configuration to maintain the communication quality.
However, if we observe at a time-slot $t_2$, there are $|y_{DL}^{(t_2)}|^2 <\gamma$ and $|y_{DL}^{(t_2-1)}|^2\geq \gamma$, we say the status has transformed from $s$ into $s+1$.
We define $t_2$ as the \textit{status transition time-slot}.
This is reasonable since the degradation of RSS (received signal strength) may indicate that the current RIS configuration can not maintain the communication quality and an update of RIS configuration is needed, which calls the beam tracking algorithm we are going to explain in the following.

In order to form our proposed beam tracking algorithm, we will propose our update rule for the RIS configuration.
Recall the optimal RIS configuration (\ref{eq: optimal conf}), we can have the optimal RIS configuration for status $s+1$ as:
\begin{equation}
(\phi_n^{(s+1)})^* = \frac{2 \pi d}{\lambda} (n-1) (\sin\theta_1-\sin\theta_2^{(s+1)}), n=1,...,N.
\label{eq: optimal conf s+1}
\end{equation}
Combined with (\ref{eq: optimal conf}), we can have the update rule of the optimal RIS configuration as:
\vspace{-0.2cm}
\begin{equation}
\phi_n^{(s+1)} = \phi_n^{(s)} - \frac{2 \pi d}{\lambda} (n-1) w^{(s+1)}, n=1,...,N,
\label{eq: update rule}
\end{equation}
where $w^{(s+1)} \triangleq \sin(\theta_2^{(s+1)})- \sin(\theta_2^{(s)})$ denotes the unknown RIS-UE channel update information.

\subsection{Time Sequential Signaling and Channel Complex Gain Model}
\label{sec: time sequential}

Previously, we have defined the status from the view of RSS.
In this subsection, we will discuss about the time sequential signaling by viewing the received signal during each time-slot.
Based on that, we will introduce our own channel complex gain update model while the UE is moving a very short distance in a very short time.

The position-dependent/time-dependent RIS-UE channel is modelled as a LOS channel decided by only two parameters, complex channel gain and angle of departure (AoD):
\begin{equation}
\mathbf{h}^{(t)} = \beta^{(t)} \mathbf{a}_{RA}(\theta_2^{(t)}),
\label{eq: h}
\end{equation}
where the superscript $t$ represents the $t$-th time-slot.
$\beta^{(t)}$ is the complex channel gain, following Rayleigh distribution.
$\mathbf{a}_{RA}(\theta_2^{(t)})$ is the steering vector and $\theta_2^{(t)}$ is the AoD.

By modelling the time-changing channel as in (\ref{eq: h}), one of the parameters, i.e., $\theta_2^{(t)}$ that describes the angle characteristic of this channel, is easy to obtain by geometrically obtaining this information.
As for the other deciding parameter $\beta^{(t)}$, we will describe our principle revealing the relationship between $\beta^{(t-1)}$ and $\beta^{(t)}$ in the following.

\begin{figure}[!t]
\centering
\vspace{-0.0 cm}
\includegraphics[width= 2 in]{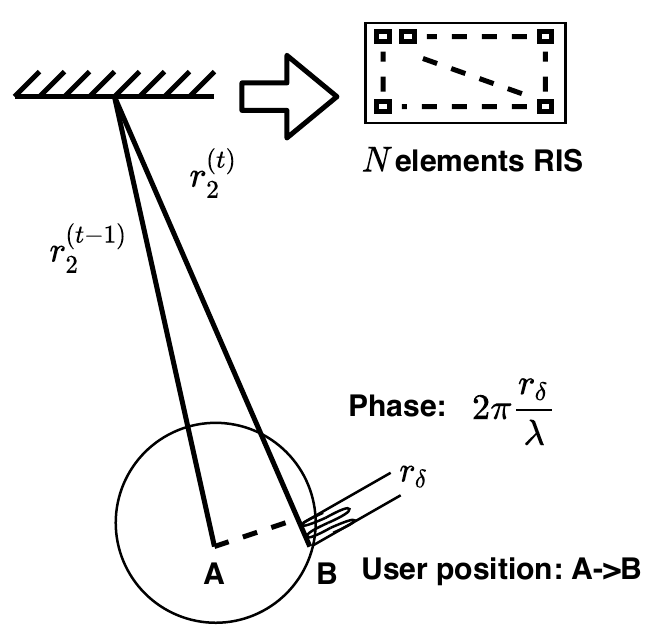}\vspace{-0.0 cm}
\caption{Complex channel gain model.}
\label{fig:complex gain}
\vspace{-0.3 cm}
\end{figure}

Since the duration of one time-slot is $t_0 = 15.6us$ according to the 11ad standard, which is very small, we make the assumption that the positions of UE at two adjacent time-slots are very near.
Therefore, the complex gain $\beta^{(t)}$ can be viewed as dependent on $\beta^{(t-1)}$ as shown in Fig. \ref{fig:complex gain}.
The relationship is revealed as follows:
\begin{equation}
\beta^{(t)} = \rho  \beta^{(t-1)}  e^{j \theta_{\delta}},
\label{eq: beta}
\end{equation}
where $\theta_{\delta} = 2 \pi \frac{r_{\delta}}{\lambda}$ describes the phase difference of the adjacent two channels caused by the difference of mmWave travel distance, i.e., $r_{\delta} = r_2^{(t)} - r_2^{(t-1)}$ and $r_2$ represents the distance between RIS and UE.
$\rho$ is a coefficient describes the path-loss relationship between these two positions.

In the following, I will introduce the path-loss-related complex channel gain update coefficient.
First, I will introduce the RSS to be used in the following explanation, i.e., $RSS^{(t)} = |y_{DL}^{(t)}|^2$.
We write out the approximation of two adjacent time-slots' received signals without considering noise and with optimal RIS configuration:
\begin{eqnarray}
y_{DL}^{(t-1)} &=& c \alpha \beta^{(t-1)} \mathbf{a}^H_{RA}(\theta_2^{(t-1)}) \mathbf{\Theta}^{(t-1)} \mathbf{a}_{RA}(\theta_1), \nonumber \\
y_{DL}^{(t)} &=& c \alpha \beta^{(t)} \mathbf{a}^H_{RA}(\theta_2^{(t)}) \mathbf{\Theta}^{(t)} \mathbf{a}_{RA}(\theta_1).
\end{eqnarray}
In this way, we can rule out the effect caused by RIS misalignment and only consider the effect of channel changed between two time-slots.
By replacing in the optimal configuration as in (\ref{eq: optimal conf}), we have the RSS for both time-slots as:
\begin{eqnarray}
|y_{DL}^{(t-1)}|^2 &=& |c \alpha \beta^{(t-1)} N|^2, \nonumber \\
|y_{DL}^{(t)}|^2 &=& |c \alpha \beta^{(t)} N|^2 = |c \alpha \beta^{(t-1)} \rho N|^2,
\label{eq: RSS equation}
\end{eqnarray}
where (\ref{eq: RSS equation}) is obtained by the definition of our channel complex gain model (\ref{eq: beta}).
Then we can have the representation of the channel complex gain coefficient $\rho$ as:
\begin{equation}
\rho^2 = \frac{ |c \alpha \beta^{(t)} N|^2 }{ |c \alpha \beta^{(t-1)} N|^2 } = \frac{|y_{DL}^{(t)}|^2}{|y_{DL}^{(t-1)}|^2} = \frac{RSS^{(t)}}{RSS^{(t-1)}}.
\end{equation}
Since the RSS is related to path-loss and having the relation of multiplication with other coefficients such as transmit antenna gain and receive antenna gain, we can have the relationship of channel complex gain coefficient $\rho$ and path-loss $PL$ as:
\begin{equation}
\rho^2 = \frac{PL^{(t-1)}}{PL^{(t)}}.
\label{eq: PL}
\end{equation}

Then let's look at the representation of path-loss \cite{PL}:
\begin{equation}
PL(r) [dB] = 20 log(\frac{4 \pi r_0}{\lambda}) + 20 log(\frac{r}{r_0}),
\end{equation}
where $r$ represents the distance between AP and UE and $r_0$ is a reference distance.
Then we know that the path-loss in multiplication relationship has the following form:
\begin{equation}
PL = (\frac{4 \pi r}{\lambda})^2.
\end{equation}
Then, according to (\ref{eq: PL}), we have the definition of channel complex gain coefficient as:
\begin{equation}
\rho = \frac{r^{(t-1)}}{r^{(t)}} = \frac{r_1 + r_2^{(t-1)}}{r_1 + r_2^{(t)}},
\label{eq: gamma}
\end{equation}
where $r_1$ is the distance between AP and RIS.
By viewing complex gain definition (\ref{eq: beta}) and equation (\ref{eq: gamma}), we find that this coefficient is related to the distance between UE and AP, which means this coefficient is determined only by the characteristic of channel itself other than the signaling procedures.
And since it is a ratio, we can extend this equation into the relationship of two adjacent status:
\begin{equation}
\rho = \sqrt{\frac{RSS^{(s+1)}}{RSS^{(s)}}} = \frac{r_1 + r_2^{(s)}}{r_1 + r_2^{(s+1)}},
\label{eq: gamma status}
\end{equation}
which is going to be used in our proposed beam tracking algorithm.

In addition, since we are representing the complex channel gain coefficient as two RSS ratio, we can always generate the time-changing RIS-UE channel $\mathbf{h}$ according to (\ref{eq: h}) after obtaining the AoD $\theta_2^{(t)}$, given the initial RIS-UE channel parameters $\beta^{(1)}$ and $\theta_2^{(1)}$, together with the update principle of the complex gain $\beta$ as in (\ref{eq: beta}).

\subsection{Fast RIS-based Beam Tracking Algorithm}

Recall the definition of different status during downlink data transmission phase in Sec. \ref{sec: state}, beam tracking algorithm takes place after status transition.
Therefore, let's focus on the transition time-slot $t_2$ as in Sec. \ref{sec: state}, where it's in status $s+1$ with RIS configuration $\Theta^{(s)}$.
We can have the approximation as:
\begin{eqnarray}
y_{DL}^{(s)} &=& c \alpha \beta^{(s)} \mathbf{a}^H_{RA}(\theta_2^{(s)}) \mathbf{\Theta}^{(s)} \mathbf{a}_{RA}(\theta_1), \nonumber \\
y_{DL}^{(t_2)} &=& c \alpha \beta^{(s+1)} \mathbf{a}^H_{RA}(\theta_2^{(s+1)}) \mathbf{\Theta}^{(s)} \mathbf{a}_{RA}(\theta_1), \label{eq: yDL t2}
\end{eqnarray}
Observing these two equations, it can be found that the channel angle information for the next status, i.e., $\theta_2^{(s+1)}$ can be represented by the parameters at status $s$, since $y_{DL}^{(t_2)}$ is utilizing $\Theta^{(s)}$ which is also utilized in $y_{DL}^{(s)}$.
As for the complex channel coefficient $\beta$, we have the relationship of two adjacent status as (\ref{eq: gamma status}),  stated in Sec. \ref{sec: time sequential}.
For simplicity and without losing generality, we use time-slot $t=1$ to represent current status $s$, and time-slot $t$ to represent next status $s+1$, where $t$ is the status transition time-slot.
And we assume that when $t=1$, the RIS has the initial optimal configuration and the RSS as:
\begin{equation}
|y_{DL}^{(1)}|^2 = |y_{DL}^{(s)}|^2 = |c \alpha \beta^{(s)} N|^2.
\label{eq: RSS yDL current}
\end{equation}
As for time-slot $t$, as the definition of status transition time-slot, we have $|y_{DL}^{(t-1)}|^2 \geq \gamma$ and $|y_{DL}^{(t)}|^2 < \gamma$.
Since the RIS configuration has not been updated as shown in (\ref{eq: yDL t2}), the decrease in RSS is caused by two factors: one is path-loss, the other is the RIS misalignment.
However, these two factors are coupled but only the information of RIS misalignment is needed for us to update the RIS configuration.
Therefore, in this subsection, we will introduce our proposed RIS beam tracking algorithm aiming at ruling out the effect of path-loss and finding the RIS optimal configuration for next status.

Rewrite the status transition time-slot received signal based on (\ref{eq: yDL t2}), by replacing $\Theta^{(s)}$ with (\ref{eq: optimal conf}), we have:
\begin{equation}
y_{DL}^{(t)} = y_{DL}^{(s+1)} = c \alpha \beta^{s+1} \sum_{n=1}^{N} e^{j [\frac{2 \pi d}{\lambda} (n-1) w^{(s+1)}]},
\label{eq: yDL s+1}
\end{equation}
and the corresponding RSS as:
\begin{equation}
|y_{DL}^{(s+1)}|^2 = |c \alpha|^2 |\beta^{(s+1)}|^2 |N^{(s+1)}|^2,
\label{eq: RSS yDL next}
\end{equation}
where we have the definition:
\begin{equation}
N^{(s+1)} \triangleq \sum_{n=1}^{N} e^{j [\frac{2 \pi d}{\lambda} (n-1) w^{(s+1)}]},
\end{equation}
where $w^{(s+1)}$ is defined in (\ref{eq: update rule}).
Since $N^{(s+1)}$ is a geometric sequence, we have:
\begin{equation}
N^{(s+1)} = \begin{cases}
                N, ~~~~~~~~ \mathrm{if} ~ w^{(s+1)} = 0, \\
                \frac{ e^{j [ \frac{2 \pi d}{\lambda} w^{(s+1)} N ] } -1 } {e^{j [ \frac{2 \pi d}{\lambda} w^{(s+1)} ] } -1}, \mathrm{if} ~ w^{(s+1)} \neq 0.
\end{cases}
\label{eq: N s+1}
\end{equation}

Given both RSSs of current status (\ref{eq: RSS yDL current}) and next status (\ref{eq: RSS yDL next}), we can have the RSS ratio as:
\begin{eqnarray}
\frac{RSS^{(s+1)}}{RSS^{(s)}} \hspace{-0.2cm}&=&\hspace{-0.2cm} \frac{|y_{DL}^{(s+1)}|^2}{|y_{DL}^{(s)}|^2} = \rho^2 \frac{|N^{(s+1)}|^2}{N^2} \nonumber \\
\hspace{-0.2cm}&=&\hspace{-0.2cm} (\frac{r^{(s)}}{r^{(s+1)}})^2 \frac{|N^{(s+1)}|^2}{N^2} = \eta^{(s+1)}.
\label{eq: ONE}
\end{eqnarray}
Observing the above equation, we can see that the drop of RSS is represented by two factors: path-loss and misalignment of RIS configuration.

In order to obtain two unknown parameters, we will need an extra equation along with (\ref{eq: ONE}).
Thus, we turn to examine the angle feature of the two received signals by observing the received signal formation as in (\ref{eq: yDL s+1}).
Similarly as we obtain the relationship of RSS ratio, we have the differential angle of two adjacent status as:
\begin{eqnarray}
 \hspace{-0.2cm}&~&\hspace{-0.2cm} \angle(y_{DL}^{(s+1)}) - \angle(y_{DL}^{(s)}) \nonumber \\
\hspace{-0.2cm}&=&\hspace{-0.2cm} \angle(\beta^{(s+1)}) + \angle(N^{(s+1)}) - \angle(\beta^{(s)}) - \angle(N^{(s)}) \nonumber \\
\hspace{-0.2cm}&=&\hspace{-0.2cm} \angle(\beta^{(s+1)}) - \angle(\beta^{(s)}) + \angle(N^{(s+1)}) \nonumber \\
\hspace{-0.2cm}&=&\hspace{-0.2cm} \theta_\delta^{(s+1)} + \angle(N^{(s+1)}) \nonumber \\
\hspace{-0.2cm}&=&\hspace{-0.2cm} \frac{2 \pi}{\lambda} (r^{(s+1)} - r^{(s)}) + \angle(N^{(s+1)}) = \xi^{(s+1)}
\label{eq: TWO}
\end{eqnarray}
According to (\ref{eq: N s+1}), we have:
\begin{equation}
\angle(N^{(s+1)}) = \begin{cases}
                0, ~~~~~~~~ \mathrm{if} ~ w^{(s+1)} = 0, \\
                \angle (\frac{ e^{j [ \frac{2 \pi d}{\lambda} w^{(s+1)} N ] } -1 } {e^{j [ \frac{2 \pi d}{\lambda} w^{(s+1)} ] } -1}) , \mathrm{if} ~ w^{(s+1)} \neq 0.
\end{cases}
\label{eq: N s+1 angle}
\end{equation}
Observing the above equation, we also see that the two uncertain parameters, i.e., path-loss and RIS misalignment, are possible to be decoupled.

Since we have the relationships of RSS ratio and differential angle of two adjacent status, the problem of finding the optimal RIS update configuration can be solved by decoupling the effects of path-loss and RIS misalignment based on (\ref{eq: ONE}) and (\ref{eq: TWO}).
Particularly, recall the update rule (\ref{eq: update rule}), we have the following optimization problem based on the fact that we considered the above deduction without noise:
\begin{equation}
\begin{split}
\{(r^{(s+1)})^*, (N^{(s+1)})^*\} = \mathrm{arg\, min} ~ \mathrm{errorI} &  + \mathrm{errorII} \\
\mathrm{s.t.} ~~~ |(\frac{r^{(s)}}{r^{(s+1)}})^2 \frac{|N^{(s+1)}|}{N} - \eta^{(s+1)}|& = \mathrm{errorI} \\
\left| \frac{2 \pi}{\lambda} |r^{(s+1)} - r^{(s)}| + \angle(N^{(s+1)}) - \xi^{(s+1)} \right|& = \mathrm{errorII}.
\end{split}
\label{eq: obj two errors}
\end{equation}


\subsection{Two-Dimensional Search-Based Solution}
\label{sec: searching}

\begin{figure}[!t]
  \centering
  \subfigure[Amplitude of $N^{(s+1)}$]{
  \includegraphics[width= 1.55 in]{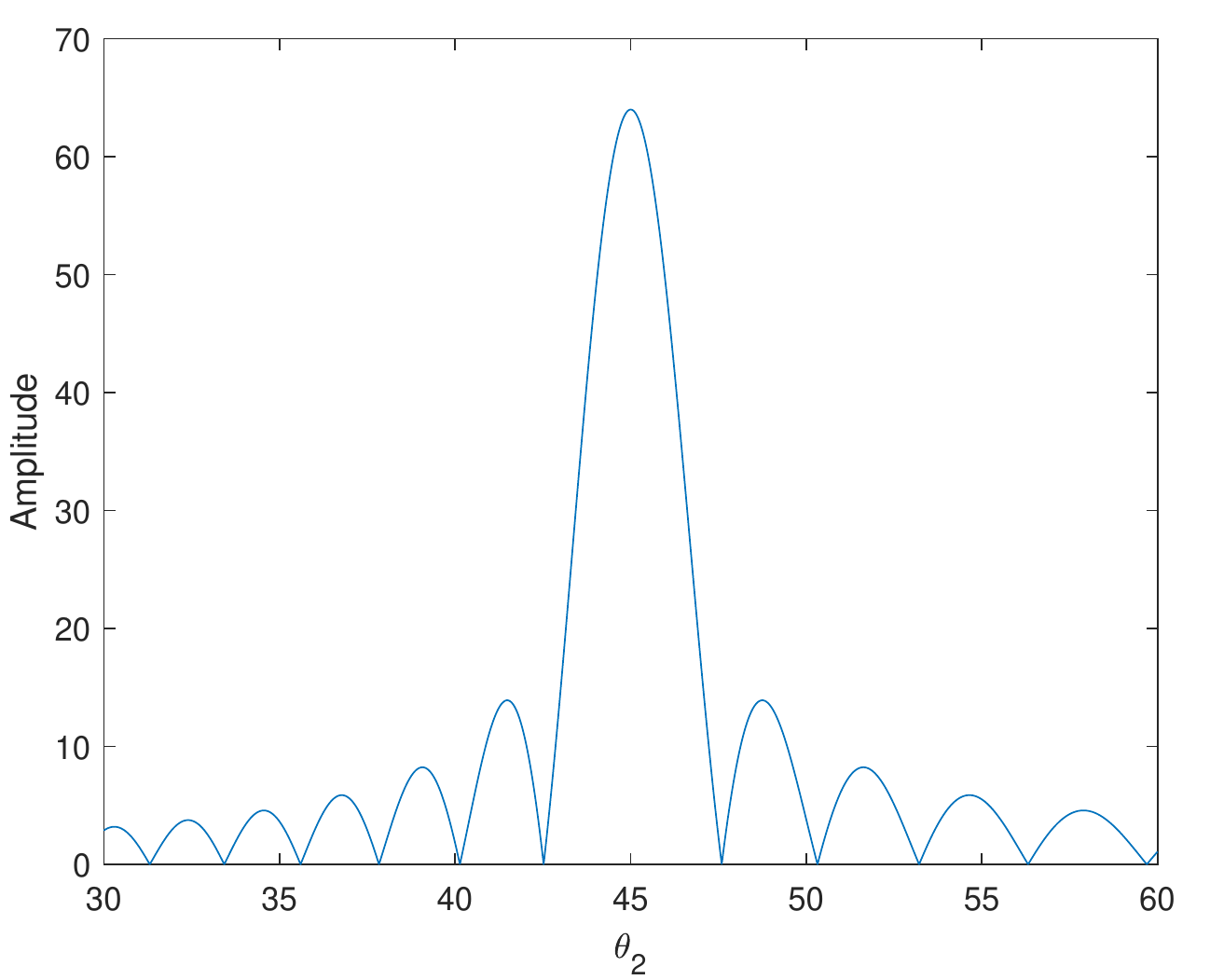}
  \label{fig:fsnorm}}
  \subfigure[Angle of $N^{(s+1)}$]{
  \includegraphics[width= 1.55 in]{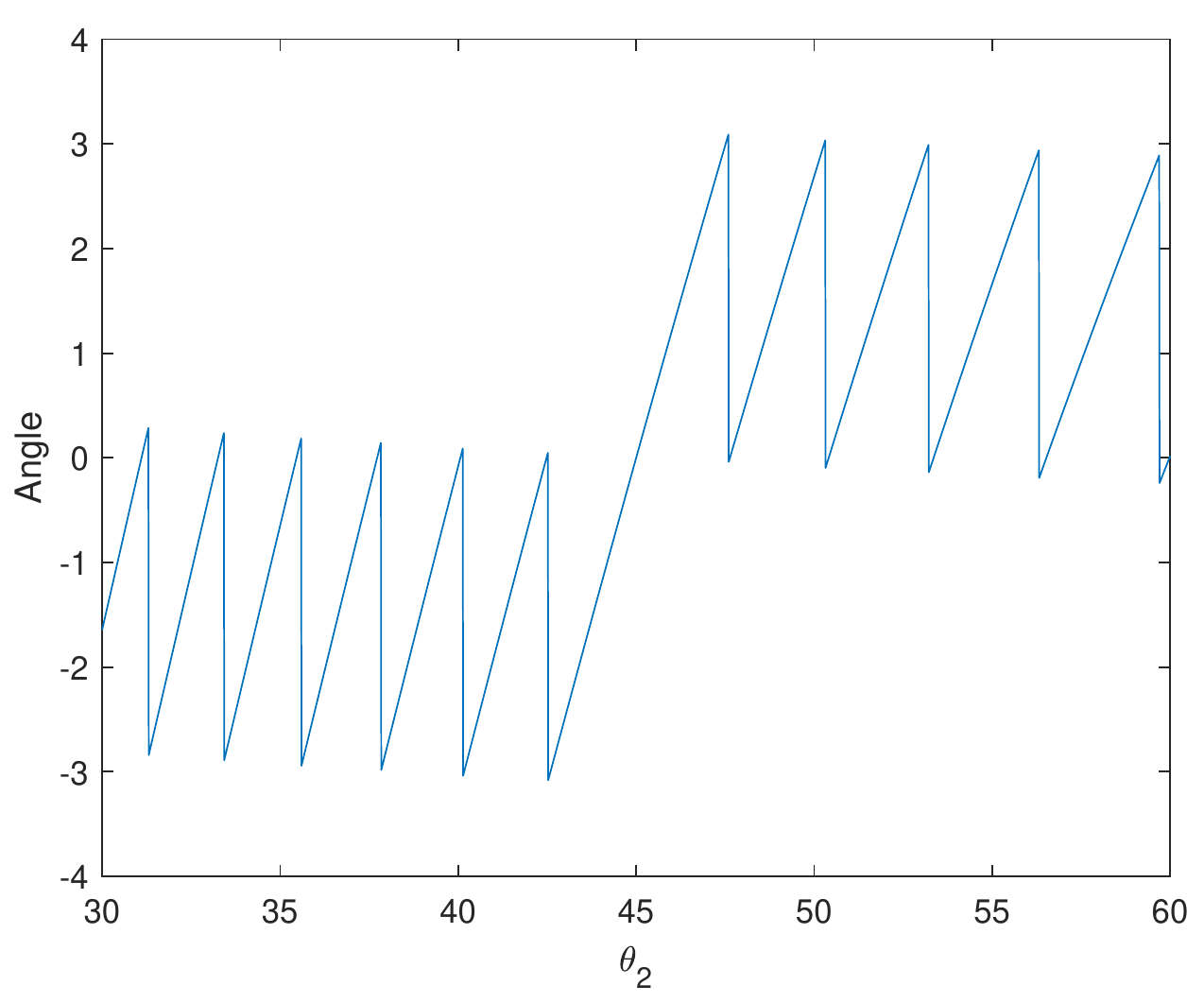}
  \label{fig:fsangle}} \vspace{-0.0cm}
  \caption{Properties of $|N^{(s+1)}|$ and $\angle (N^{(s+1)}$).}\label{fig:fs} \vspace{-0.0cm}
\end{figure}

In this subsection, we will illustrate the solution of the optimization problem (\ref{eq: obj two errors}).
We propose a searching based solution by calculating both errors of possible uncertain parameter pairs, i.e., errorI and errorII.
Such idea is inspired by the observation of the characteristics of $|N^{(s+1)}|^2$ and $\angle(N^{(s+1)})$ as shown in Fig. \ref{fig:fs}.
The above figures are obtained when we set $\theta_1 = 45 ^\circ$.
It can be seen that as long as we keep the searching angle of $\theta_2$ within a certain range, we can always obtain a rather large outcome of $|N^{(s+1)}|^2$.
Also, the angle characteristic of it can be viewed as linear within the same $\theta_2$ changing range.
Based on these observations and the intuition that the AoD changing happens within a certain range in respect to the optimal $\theta_2$ at time sequence $t=1$, i.e., when RIS configuration is optimal at status $s$, we have our search-based solution to problem (\ref{eq: obj two errors}) as described below.

%
%

We first set the $\theta_2^{(s+1)}$ searching range as $\pm 2.5 ^\circ$ centered at the optimal $\theta_2^{(s)}$, and calculate the corresponding $N^{(s+1)}$ according to (\ref{eq: N s+1}).
In this case, the center of $|N^{(s+1)}|^2$ is when $w^{(s+1)} = 0$, i.e., we start the searching procedure from status $s$ when RIS configuration is optimal.
Then according to the first constraint equation in (\ref{eq: obj two errors}), and the calculated series of $N^{(s+1)}$, we can calculate the corresponding $r_{cal}^{(s+1)}$ as:
\begin{equation}
r_{cal}^{(s+1)} = \frac{r^{(s)} |N^{(s+1)}|}{\sqrt{\eta^{(s+1)}} N}.
\label{eq: r cal}
\end{equation}
Next, we set the $r^{(s+1)}$ searching range as $\pm 0.005m$ centered at the calculated series of $r_{cal}^{(s+1)}$.
With multiple $N^{(s+1)}$ and $r^{(s+1)}$ pairs, we can calculate the summation of the corresponding two errors according to (\ref{eq: obj two errors}).
Then we select the best several pairs of $N^{(s+1)}$ and $r^{(s+1)}$ as the best several solutions to problem (\ref{eq: obj two errors}).
The algorithm for solving this problem is illustrated in Algorithm \ref{agsearching}.

\begin{algorithm}[t!]
\caption {Two-Dimensional Search-Based Solution}
\label{agsearching}
\begin{algorithmic}[1]
\REQUIRE A series of $\theta_2^{(s+1)}$ centered at $\theta_2^{(s)}$ with searching range as $\pm 2.5^\circ$.\\
\ENSURE Best $N_{sol}$ pairs of $N^{(s+1)}$ and $r^{(s+1)}$.\\
\STATE Calculate corresponding series of $N^{(s+1)}$ according to (\ref{eq: N s+1}) based on a series of $\theta_2^{(s+1)}$.
\STATE Calculate corresponding series of $r_{cal}^{(s+1)}$ according to (\ref{eq: r cal}).
\STATE \textbf{for} each $r_{cal}^{(s+1)}$ \textbf{do}
\STATE  ~~ Obtain series of $r^{(s+1)}$ centered at $r_{cal}^{(s+1)}$ with searching range as $\pm 0.005m$.\\
 \textbf{end for}
\STATE \textbf{for} each pair of $N^{(s+1)}$ and $r^{(s+1)}$ \textbf{do}
\STATE  ~~ Calculate $\mathrm{errorI} + \mathrm{errorII}$ in (\ref{eq: obj two errors}).\\
 \textbf{end for}
\STATE Select best $N_{sol}$ pairs of $N^{(s+1)}$ and $r^{(s+1)}$ that has minimum error summation.
\end{algorithmic}
\end{algorithm}

\subsection{Overall Signaling Procedure and Framework}
\label{sec: overall}

\begin{figure}[t]
\centering
\vspace{-0.0 cm}
\includegraphics[width= 3.5 in]{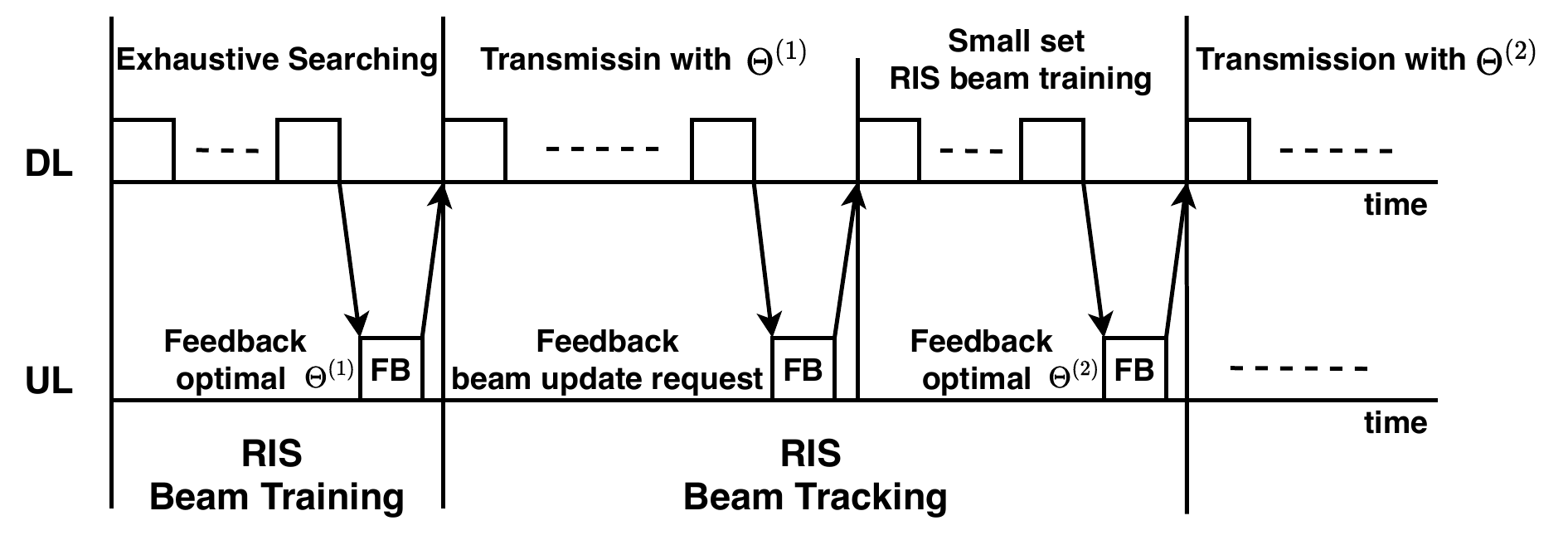}\vspace{-0.0 cm}
\caption{The overall signaling procedure with our proposed fast beam tracking algorithm.}
\label{fig: protocol}\vspace{-0.0cm}
\end{figure}

\begin{figure}[t]
\centering
\vspace{-0.5 cm}
\includegraphics[width= 3 in]{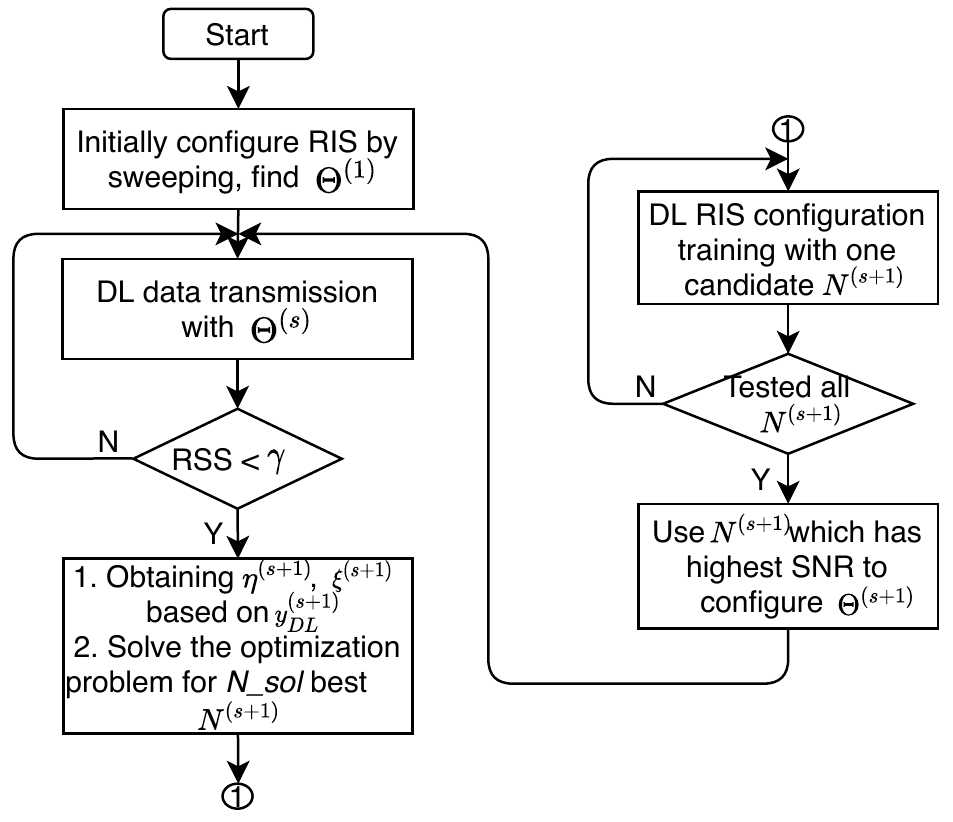}\vspace{-0.0 cm}
\caption{The framework of our proposed beam tracking algorithm.}
\label{fig: framework}\vspace{0.0cm}
\end{figure}

To clearly demonstrate the whole working process of our proposed RIS-based fast beam tracking algorithm, we illustrate the overall signaling procedure between a fixed RIS and a mobile UE, including both RIS beam training phase and RIS beam tracking phase in Fig. \ref{fig: protocol}.
We also illustrate the framework of the whole process in Fig. \ref{fig: framework}.

Initially, the configuration of RIS is done by sweeping the all possible phases, which corresponds to the RIS beam training phase as in Fig. \ref{fig: protocol}.
This phase takes the exhaustive searching strategy thus will cost a large amount of signaling time.
After obtaining the optimal initial RIS configuration $\Theta^{(1)}$, it goes into downlink data transmission phase where the RIS beam tracking algorithm is working.
During such phase, the user will observe the RSS at each time-slot and compare it with a certain threshold $\gamma$.
Once the RSS is below the threshold, the user will feedback a signal indicating that RIS configuration update is needed.
At the same time, the two parameters of RSS ratios and received signal differential angle are also fedback to AP in order to solve for the optimization problem according to Algorithm \ref{agsearching}.
Notice that such complex procedure is done by calculating at AP side, which saves a large amount of signaling time.
After solving for $N_{sol}$ best possible solutions $N^{(s+1)}$ for updating RIS configuration, a small set downlink training is processed.
Notice that such time is rather short since the complexity is transformed into the calculation complexity happened at AP.
After testing all candidate solutions, the one which can has the highest SNR will be utilized as the update information to configure $\Theta^{(s+1)}$.

\section{Simulation Evaluations}
\label{sec: sim}

In this section, we will first describe how to generate the ground true channel with user mobility according to our channel complex gain model as in Sec. \ref{sec: time sequential} followed by extensive evaluations based on it.

\subsection{User Trajectory Generation and Simulation Parameters}
\label{sec: user}

\begin{figure}[!t]
\centering
\vspace{-0.0 cm}
\includegraphics[width= 2.5 in]{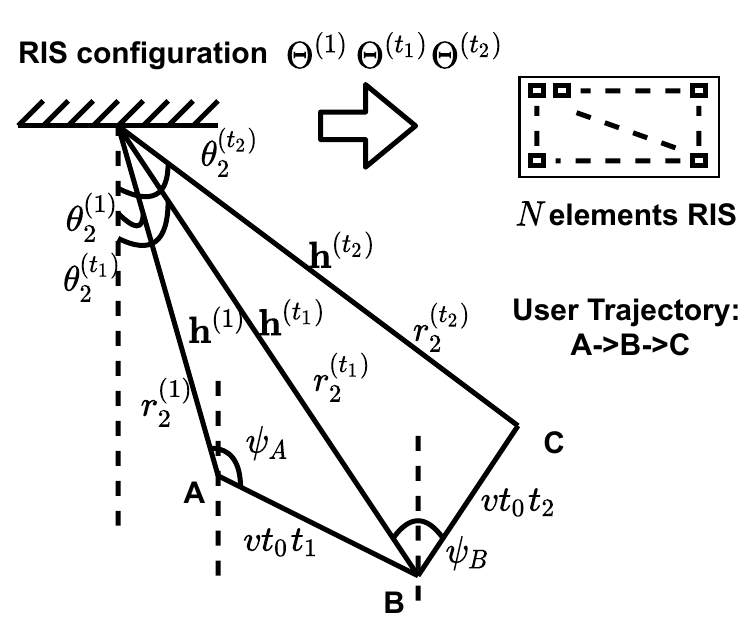}\vspace{-0.0 cm}
\caption{User trajectory in simulation study.}
\label{fig:simulationpath}
\vspace{-0.3 cm}
\end{figure}

In this subsection, we will showcase how to generate the user trajectory and the corresponding parameters in our simulation studies as shown in Fig. \ref{fig:simulationpath}.
We generate our ground true channel as described below:

Firstly, we want to generate the ground true channel for path $AB$ as illustrated in Fig. \ref{fig:simulationpath}, where a human is walking along a straight line starting from point A to point B.
To decide the channels for path $AB$, we only need to know the starting position, the direction of walking and the distance of path $AB$.
We decide the starting point A with initial AoD at RIS set as $\theta_2^{(1)}$ and the initial distance between RIS and UE as $r_2^{(1)}$.
Then we assume the angle between RIS-to-A and path $AB$ is set as $\psi_A$, which indicates the direction the user is walking.
The time duration of walking along path $AB$ is set to $t_1$, indicating the distance of path $|AB|=vt_0t_1$.

There are four steps to generate the ground true channel $\mathbf{h}$ for path $AB$.
\begin{itemize}
    \item \textbf{Step 1}: Decide the position of starting point A.
    In this step, we initialize A by assigning $r_2^{(1)}$ and $\psi_A = 110^\circ$, with the initial AoD as $\theta_2^{(1)} = 20 ^\circ$, meaning that path $AB$ is parallel to the RIS.
    
    \item \textbf{Step 2}: Decide the number of time-slots when UE moves along path $AB$.
    In this step, we calculate the number of time-slots $n$.
    We first calculate the distance UE moves during one time-slot $t_0 = 15.6us$ by $vt_0$, where $v$ represents the speed of UE.
    Then we assign the distance of path $AB$ as $|AB|$.
    The number of time-slots on this path is calculated by $n = \lceil |AB|/(vt_0) \rceil$, then $t = 1:n$ meaning each transmission time-slot.
    
    \item \textbf{Step 3}: Decide the complex channel gain.
    The complex channel gain $\beta^{(1)}$ is generated following Rayleigh distribution.
    
    \item \textbf{Step 4}: Generate the channel $\mathbf{h}$ along path $AB$ according to geometrical relationship.
    Given the distance of AP-RIS as $r_1$, the update rule of RIS-UE distance $r_2$ is as follows:
    \begin{equation}
    r_2^{(t)} = \sqrt{ r_1^2 + (vt*t_0) - 2 r_1 (vt*t_0) cos(\psi_A) }.
    \end{equation}
    Then following the update rule of complex gain (\ref{eq: beta}) and coefficient (\ref{eq: gamma}), we can get the complex channel gain for each time-slot.
    To update the AoD $\theta_2$, we use the equation below:
    \begin{equation}
    cos(\theta_2^{(t)} - \theta_2^{(1)}) = \frac{r_1^2 + (r_2^{(t)})^2 - (vt*t_0)^2}{2 r_1 r_2^{(t)}}.
    \end{equation}
    Then the channel at time-slot $t$ can be obtained according to (\ref{eq: h}).
\end{itemize}

When the UE is changing direction during the walk, the same procedure can be done to generate the channel for the new path, e.g., path $BC$ as is shown in Fig. \ref{fig:simulationpath}.
Specifically, we need the position of starting point $B$, which is also the ending point of path $AB$ and such information can be easily acquired.

In the following subsections, we will illustrate the simulation results of the proposed RIS configuration algorithm under node mobility in mmWave systems based on the channel generated in this subsection.

Consider a mmWave system including an AP equipped with $N_t = 16$ Uniform Linear Antennas (ULA), whose spacing is $d=\frac{\lambda}{2}$, a single antenna mobile UE, and a RIS equipped with $N = 64$ elements, which is able to achieve continuous phase shifting.
The AoA at RIS side is assumed to be $\theta_1 = 45^\circ$.
For simplicity, the noise variances are set to 1 and transmission power is SNR=10dB.
We will be comparing three algorithms in the simulations.
First is our proposed algorithm, where the search resolutions for $N^{(s+1)}$ and $r^{(s+1)}$ is set as $\pm 2.5^\circ$ and $\pm 0.005m$.
The threshold while calling RIS beam tracking algorithm is set as $\gamma = 0.9$ except for specifically assignment.
The number of downlink RIS beam training time frames is set as $7$, which is a reasonable choice as can be explained in Fig. \ref{fig: fig5}.
Second is the exhaustive searching idea, where the number of downlink RIS beam training time frames is decided by the resolution of RIS PSs.
Since the procedure of beam searching is called after communication outage happens, we set the threshold for this idea as $\gamma_{exh}=0.5$.
Third is the oracle case, where RIS will always find the optimal configuration after the RSS is below the threshold of $\gamma = 0.9$ in order to compare with our proposed algorithm.
In other words, the oracle idea will not have downlink RIS beam training time frames.

\subsection{Effect of Noise on Instantaneous Rate}
\label{sec:rate}

\begin{figure}[!t]
  \centering
  \subfigure[Proposed algorithm.]{
  \includegraphics[width= 1.55 in]{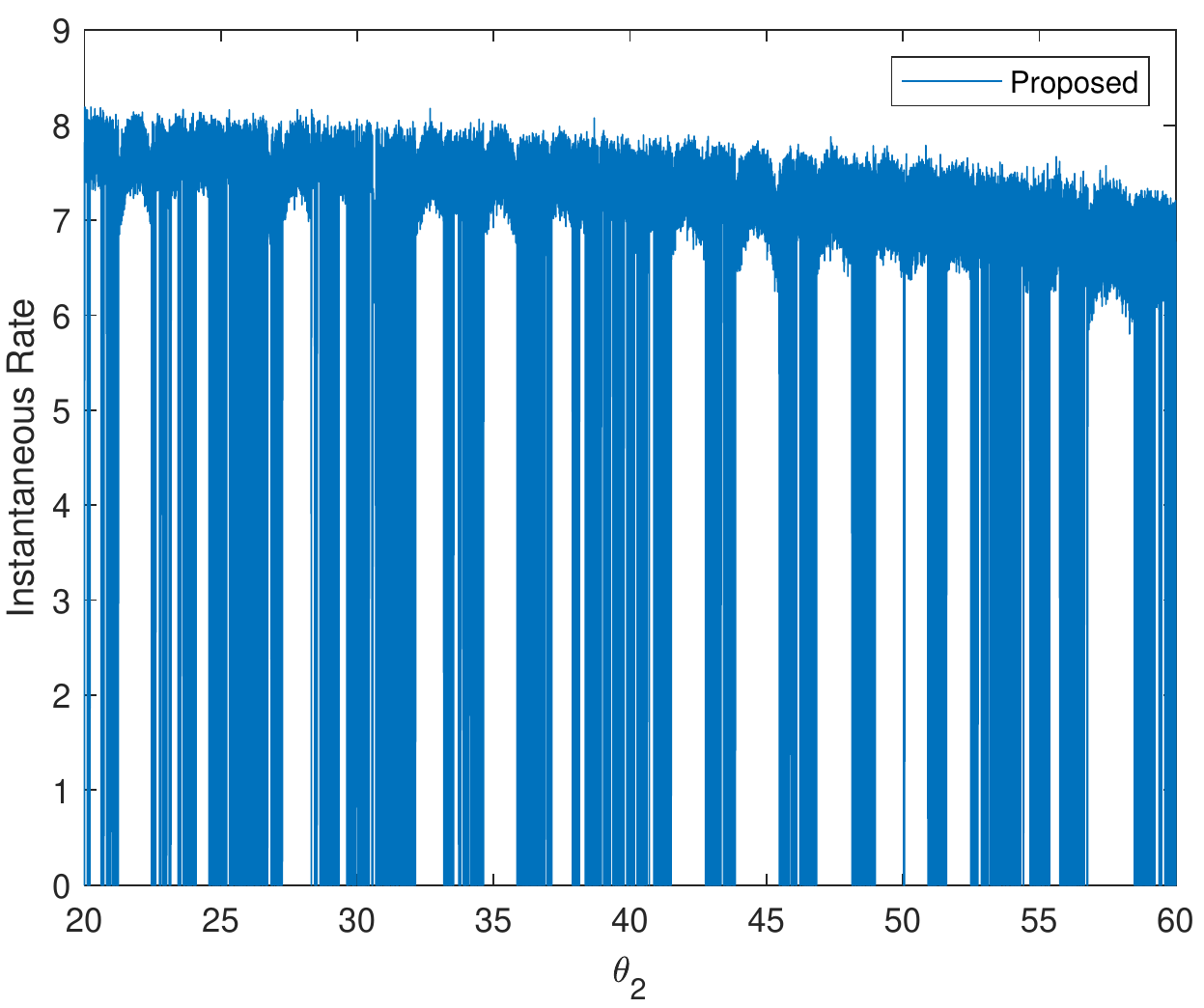}
  \label{fig:fig21}}
  \subfigure[Exhaustive search, res = $1^\circ$.]{
  \includegraphics[width= 1.55 in]{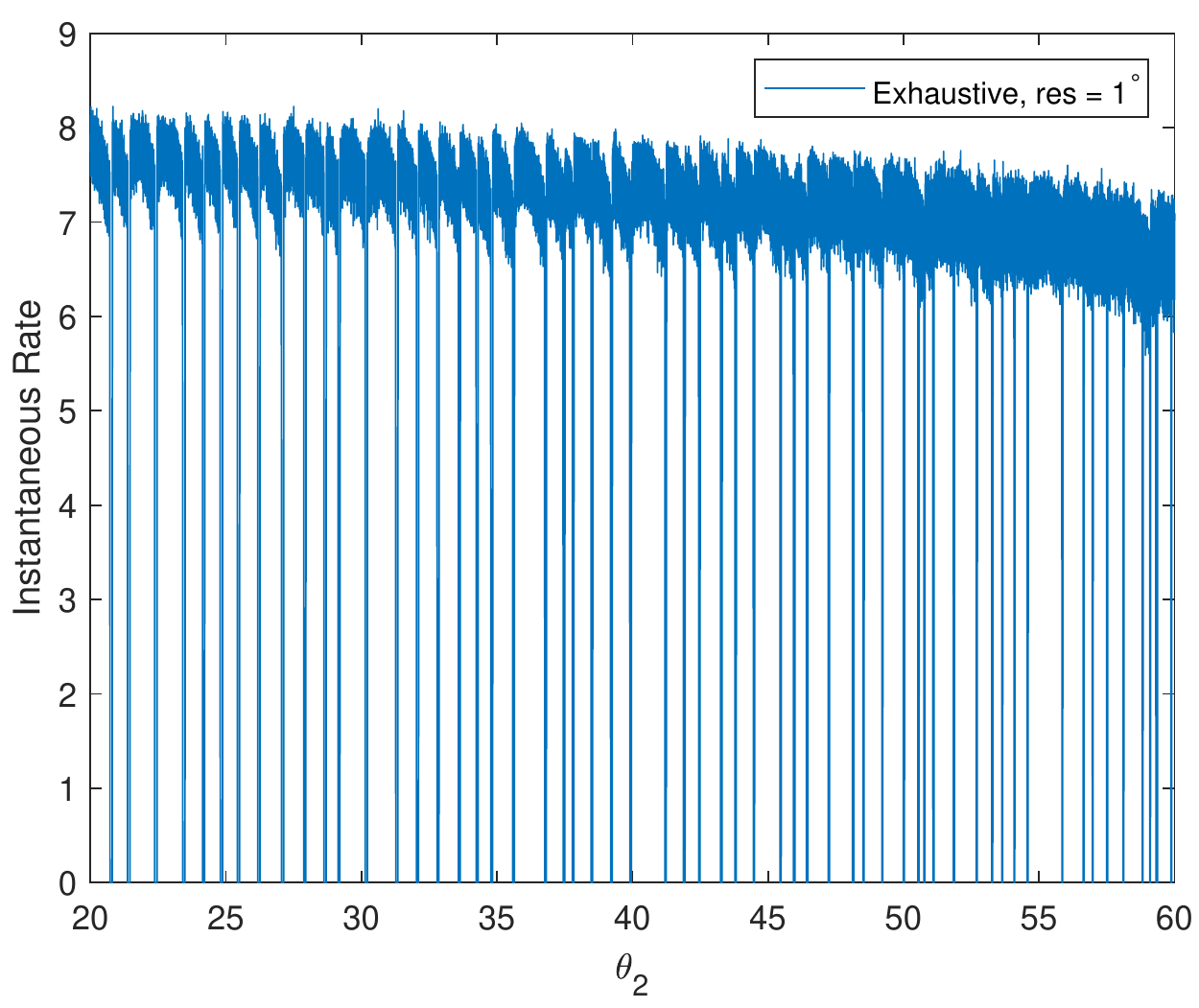}
  \label{fig:fig22}} \vspace{-0.0cm}
  \subfigure[Exhaustive search, res = $5^\circ$.]{
  \includegraphics[width= 1.55 in]{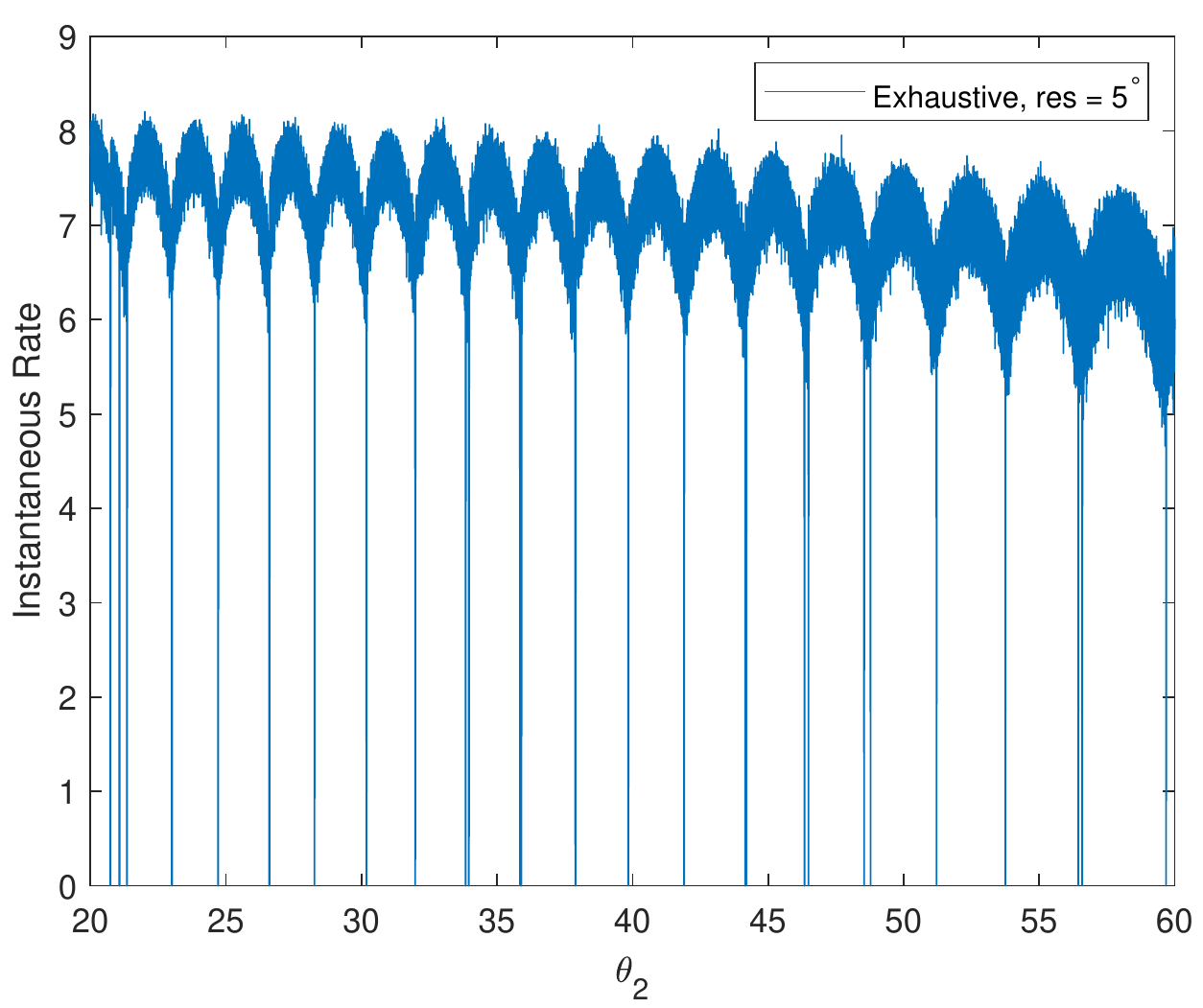}
  \label{fig:fig23}}
  \subfigure[Exhaustive search, res = $10^\circ$.]{
  \includegraphics[width= 1.55 in]{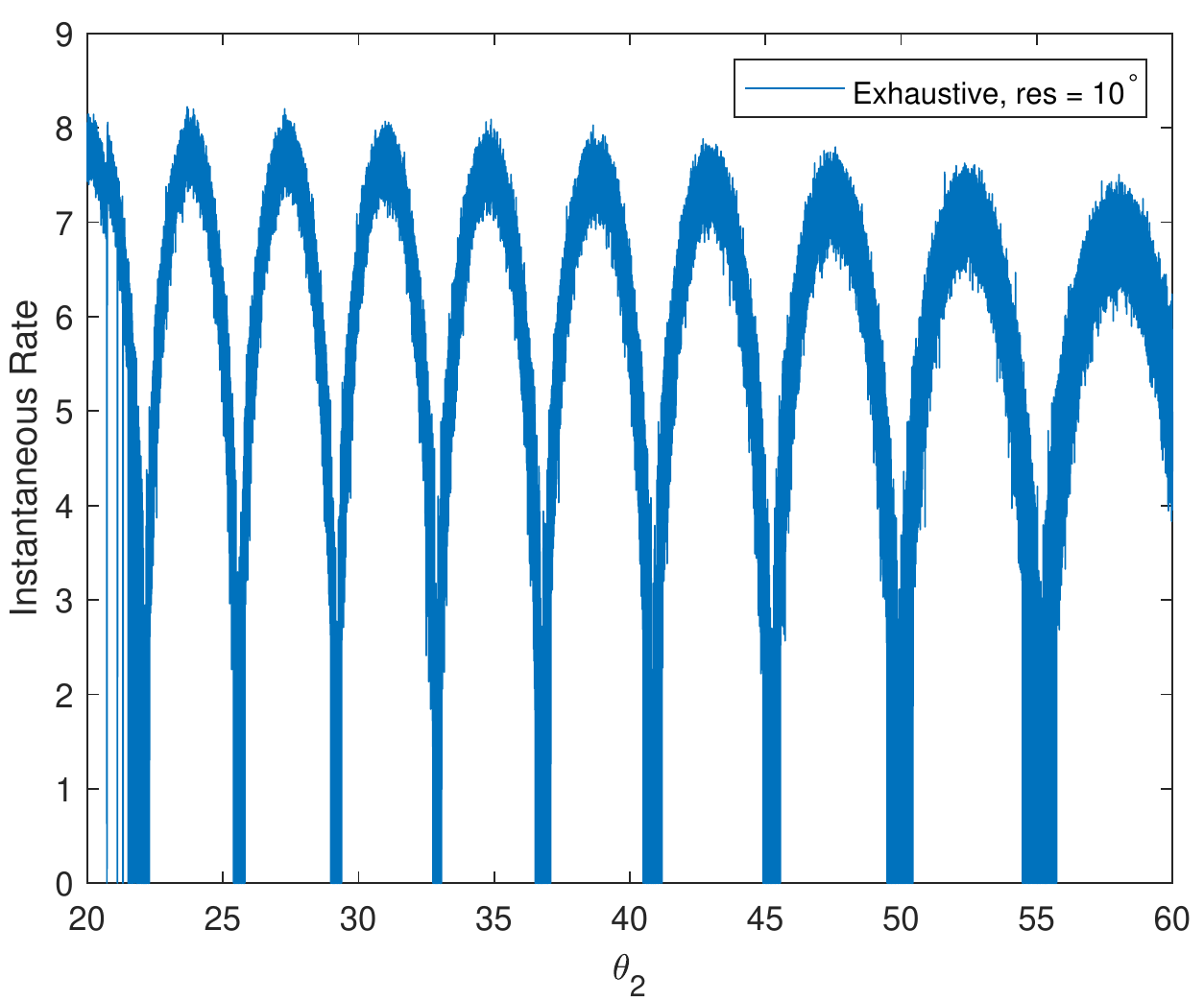}
  \label{fig:fig24}}
  \caption{Instantaneous rate (bit/s/Hz) versus user position in spatial angle $\theta_2$ (degree) (Distance of AP-RIS $r_1=4m$, distance of RIS-UE $r_2=4m$, user walking speed $v=0.6m/s$).}
  \label{fig:fig2} \vspace{-0.0cm}
\end{figure}

We first look at the performance of instantaneous rate comparing both our proposed algorithm and the exhaustive searching strategy in Figs. \ref{fig:fig2}.
In our simulation, we consider the instantaneous rate as 0 when DL data transmission is not happening.
Such time-slots including DL signaling time and UL feedback time.
From the figure, we can see that when considering noise, the instantaneous rates are fluctuating with their trends clear.
By observing such trend, we can see that both our proposed algorithm and the exhaustive search idea with $1^\circ$ resolution can obtain near-optimal RIS configuration almost every time-slot, thus maintain the rate at a rather high level most of the time.
The fact that coarser resolution will deteriorate instantaneous rate can be verified by comparing Figs. \ref{fig:fig22}-\ref{fig:fig24}.
However, at the same time, finer resolution will induce higher signaling time during DL training phase.
Therefore, there's a trade-off between instantaneous rate and signaling time.
The idea of our proposed beam tracking algorithm is to shorten the signaling time while maintaining the instantaneous rate above a certain threshold.

\subsection{Cumulative Average Rate}
\label{sec:cumulative}

\begin{figure}[t]
\centering
\vspace{-0.0 cm}
\includegraphics[width= 3.3 in]{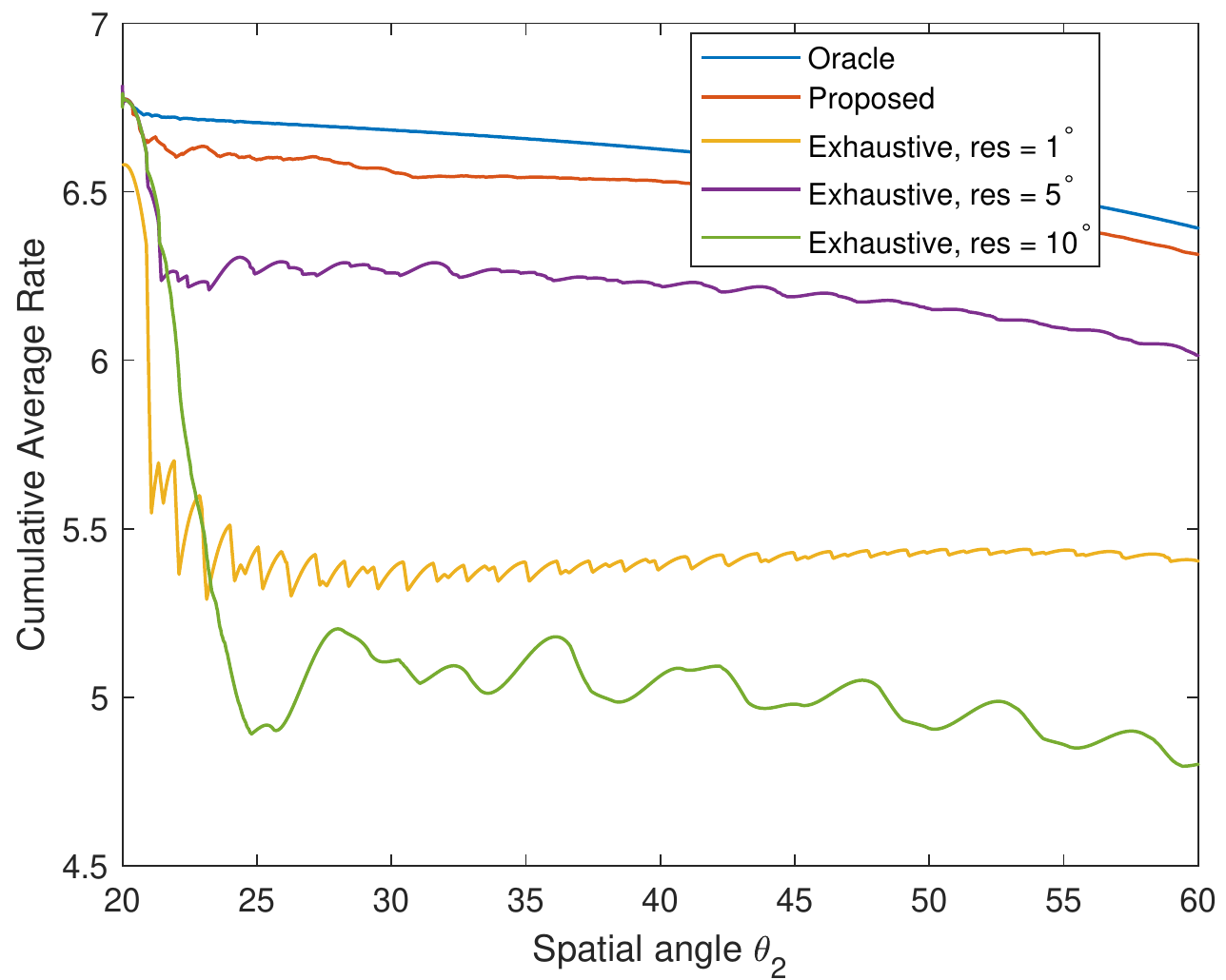}\vspace{-0.0 cm}
\caption{Cumulative average rate (bit/Hz) versus user position in spatial angle $\theta_2$ (degree) (Number of transmit antennas $N_t=16$, number of RIS elements $N=64$, SNR = 10dB, proposed algorithm threshold $\gamma = 0.9$, exhaustive searching idea threshold $\gamma_{exh}=0.5$, exhaustive searching resolution $1^\circ$, $5^\circ$, $10^\circ$, and walking speed $v=0.6m/s$).}
\label{fig:fig1}\vspace{-0.00 cm}
\end{figure}

In order to evaluate the overall performance of the three RIS configuration ideas, we propose to use cumulative average rate to reflect the trade-off of the signaling time and instantaneous rate.
The cumulative average rate is calculated by the summation of instantaneous rates over total number of time-slots and has the following update rule:
\begin{equation}
    R_{cum}^{(t+1)} = \frac{\sum_{i=1}^{t+1} R_{ins}^{(i)}}{t+1} = \frac{t R_{cum}^{(t)} + R_{ins}^{(t+1)}}{t+1},
\end{equation}
where $R_{cum}$ stands for the cumulative average rate and $R_{ins}$ stands for the instantaneous rate.
The superscript stands for the corresponding time-slot.

We compare our proposed beam tracking algorithm with the oracle one, serving as the upper bound together with the exhaustive searching strategy with three different resolution settings in Fig. \ref{fig:fig1}.
Different from Figs. \ref{fig:fig2}, which reflect instantaneous rate and will always have certain positions that can reach the highest achievable instantaneous rate, cumulative average rates for different algorithms has gaps.
It can be seen that the oracle idea has the highest performance since it has no cost on signaling time.
Our proposed beam tracking algorithm has near-optimal performance since we tried to transform the time cost into calculations happen inside the BS, significantly saving signaling time comparing with conventional exhaustive search ideas.
In addition, we found that the performance of exhaustive search idea with $5^\circ$ resolution outperforms when with $1^\circ$ and $10^\circ$ resolutions.
This is because that with finer resolution, the accuracy of the updated RIS configuration is higher, which explains why it outperforms when res = $10^\circ$.
However, with finer resolution, the signaling time for each search will get longer, resulting in lower cumulative average rate since the number of time-slots when $R_{ins}^{(t)}=0$ gets higher, which explains why it outperforms when res = $1^\circ$.
According to Fig. \ref{fig:fig1}, cumulative average rate can effectively reflect the trade-off of signaling time and instantaneous rate and is suitable when evaluting the overall performance.


\subsection{Effect of User Velocity}
\label{sec:velocity}

\begin{figure}[t]
\centering
\includegraphics[width= 3.3 in]{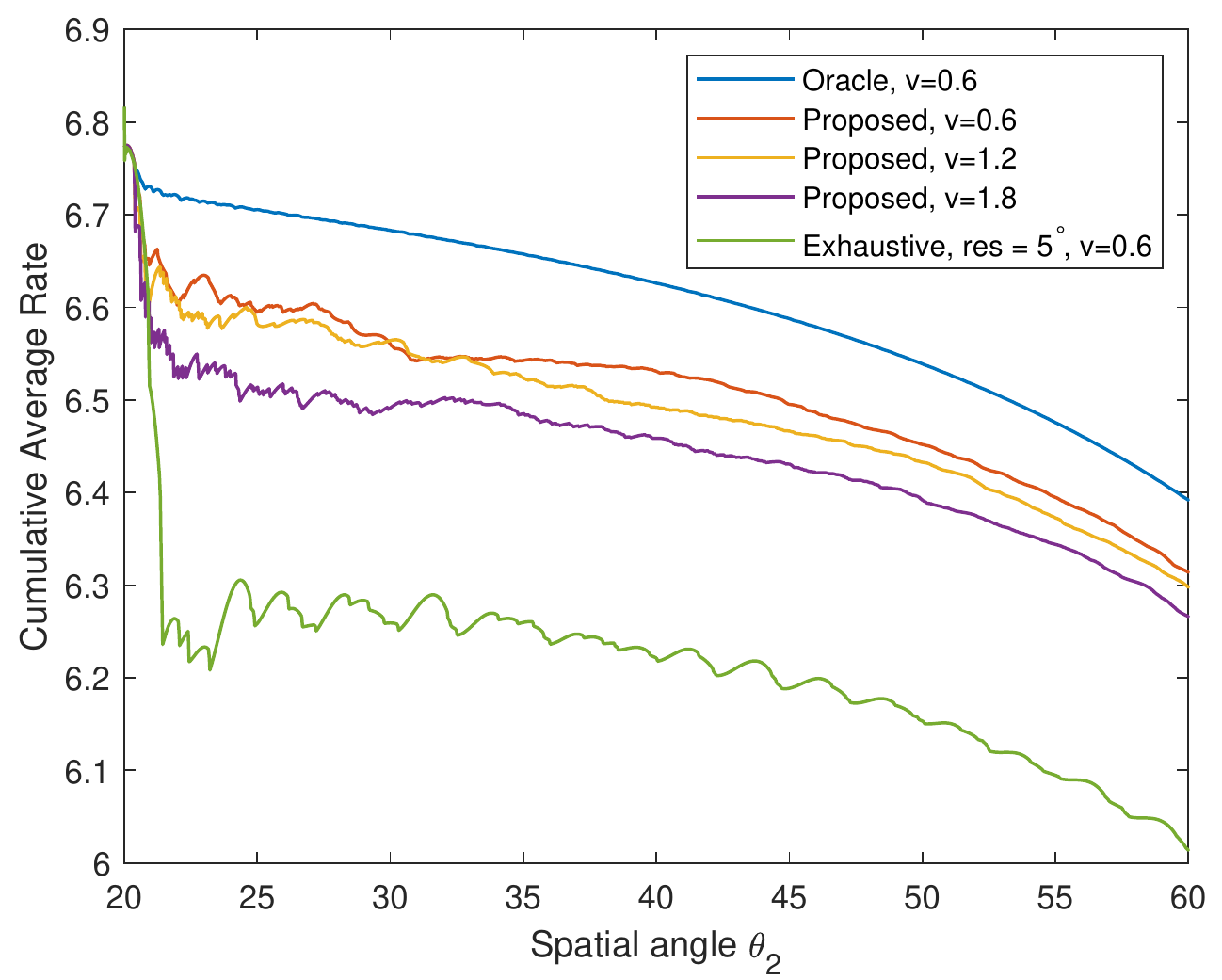}\vspace{-0.0 cm}
\caption{Cumulative average rate (bit/Hz) versus user position in spatial angle $\theta_2$ (degree) (Number of transmit antennas $N_t=16$, number of RIS elements $N=64$, SNR = 10dB, proposed algorithm threshold $\gamma = 0.9$, exhaustive searching idea threshold $\gamma_{exh}=0.5$, and walking speed $v=0.6m/s$, $v=1.2m/s$ and $v=1.8m/s$).}
\label{fig: fig3}\vspace{-0.0 cm}
\end{figure}

Then we turn to evaluate the effect of user velocity on our proposed beam tracking algorithm. 
We compare our algorithm with velocities as $v=0.6m/s$, $v=1.2m/s$, $v=1.8m/s$, respectively, along with the upper bound and the conventional exhaustive searching idea with $5^\circ$ resolution.
It can be seen from Fig. \ref{fig: fig3} that with lower user velocity, the proposed algorithm has higher cumulative average rate.
Intuitively, with lower user velocity, the angle changed during a same amount of time will be smaller, meaning that one particular configuration of RIS will have longer time having the communication quality above the same threshold.
Therefore, the frequency of implementing the beam tracking algorithm will be lower, thus can save signaling time and achieves higher cumulative average rate.

\subsection{Effect of Threshold}
\label{sec:threshold}

\begin{figure}[t]
\centering
\includegraphics[width= 3.3 in]{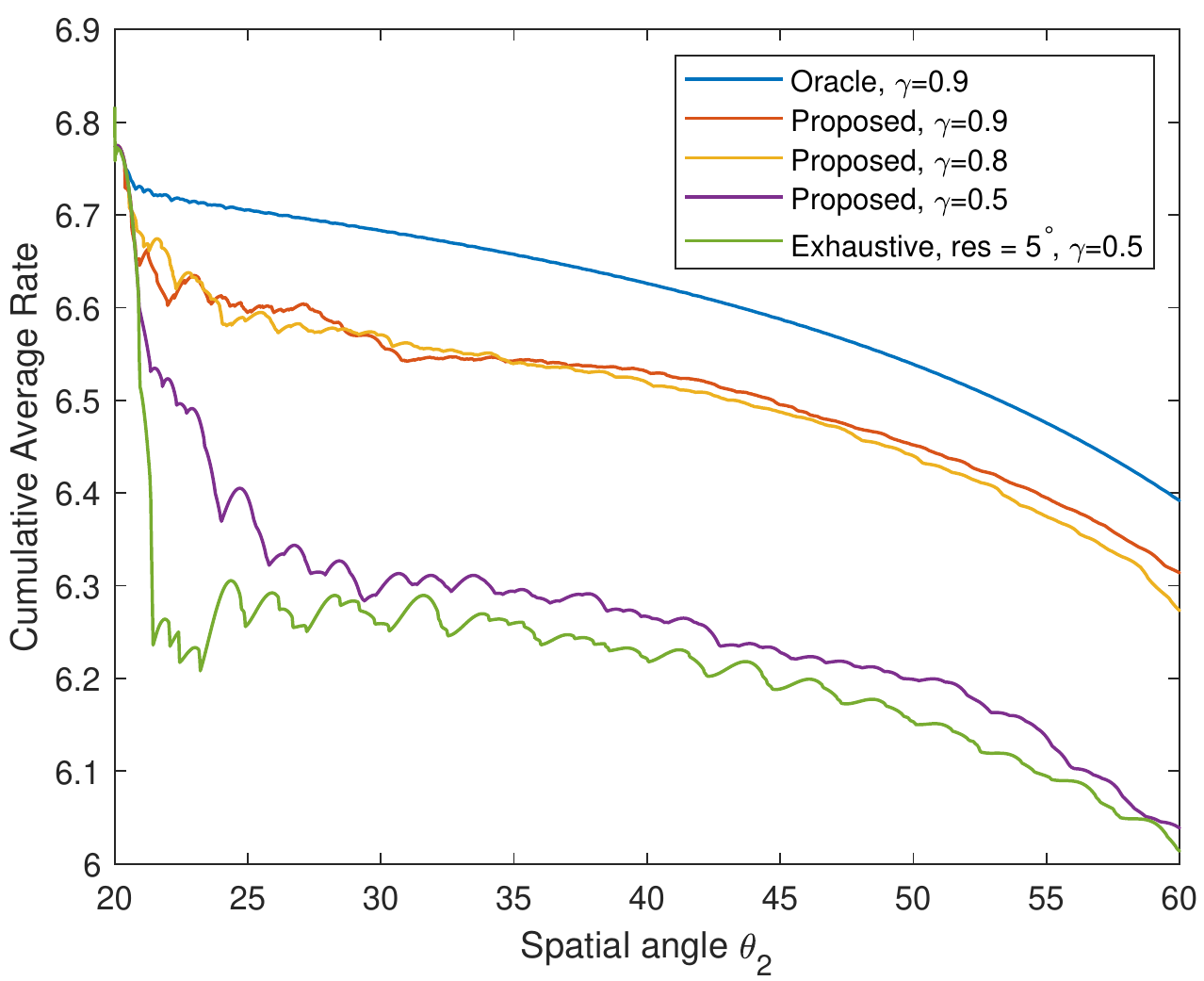}\vspace{-0.0 cm}
\caption{Cumulative average rate (bit/Hz) versus user position in spatial angle $\theta_2$ (degree) (Number of transmit antennas $N_t=16$, number of RIS elements $N=64$, SNR = 10dB, proposed algorithm threshold $\gamma = 0.9$, $\gamma = 0.8$, $\gamma = 0.5$, exhaustive searching idea threshold $\gamma_{exh}=0.5$, and walking speed $v=0.6m/s$).}
\label{fig: fig4}\vspace{-0.0 cm}
\end{figure}

Next we examine the effect of threshold on our proposed beam tracking algorithm.
We compare our algorithm with thresholds set as $\gamma=0.9$, $\gamma=0.8$ and $\gamma=0.5$, respectively, along with upper bound and the exhaustive searching idea with $5^\circ$ resolution.
As is shown in Fig. \ref{fig: fig4}, with higher threshold, which means higher communication quality demand, the performance of cumulative average rate will be higher.
Comparing with the exhaustive strategy with the same threshold when $\gamma=0.5$, our proposed algorithm still outperforms it, which verifies that our proposed beam tracking algorithm saves much more signaling time than the conventional idea.

\subsection{Effect of Candidate Set Size}
\label{sec:searching range}

The signaling time of our proposed algorithm depends on the size of candidate solution for the optimization problem (\ref{eq: obj two errors}).
Intuitively, with larger size of the candidate set, the chance of selecting the suitable RIS configuration is higher at the cost of occupying longer signaling time.
However, if the size of such set is too small, it is possible that the solution for (\ref{eq: obj two errors}) will not include the suitable RIS configuration.
Therefore, keeping the size of such set as small as possible while maintaining a rather small rate difference with the oracle idea will get us higher cumulative average rate.



\begin{figure}[t]
\centering
\includegraphics[width= 3.3 in]{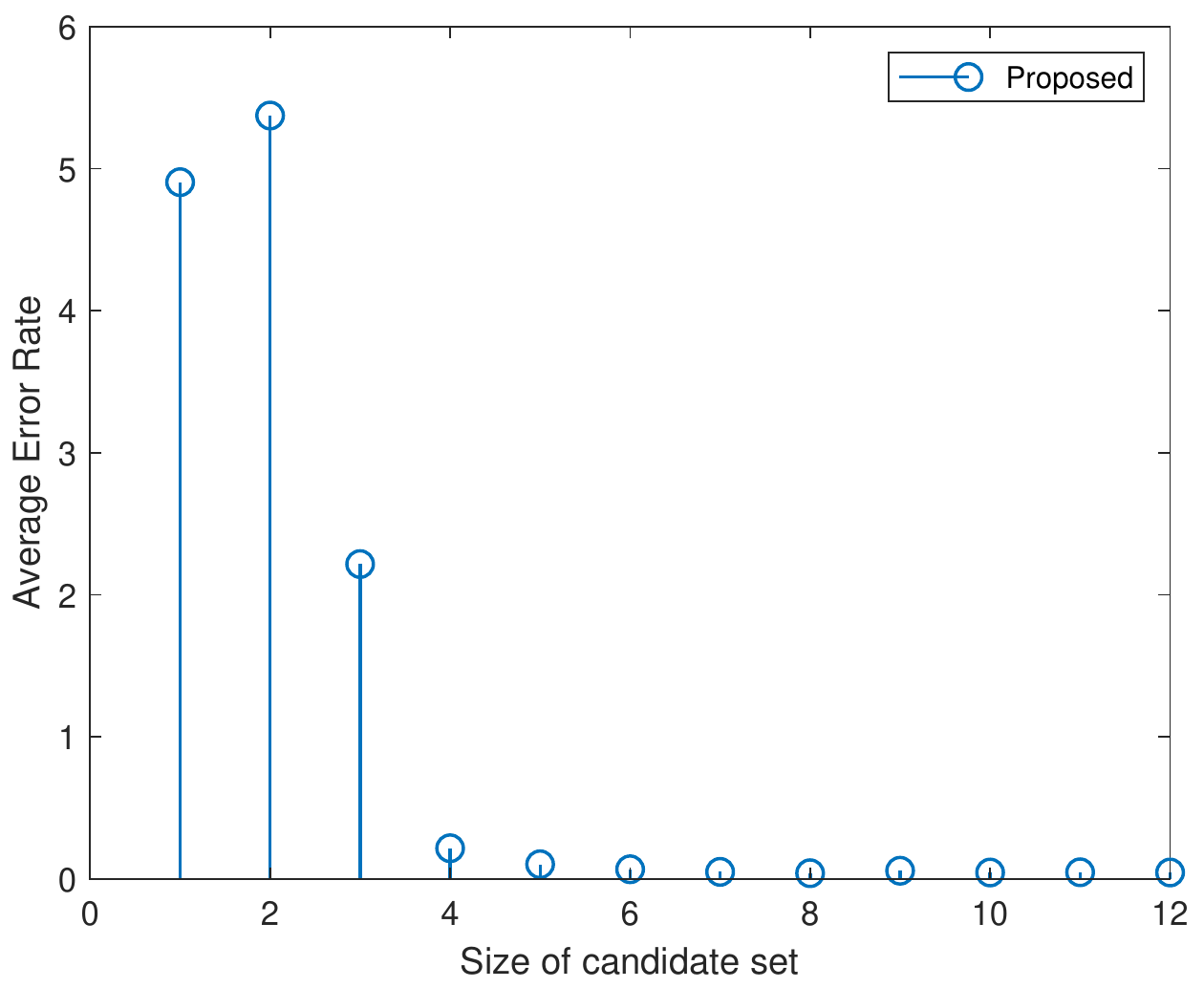}\vspace{-0.0 cm}
\caption{Average error rate (bit/s/Hz) versus size of candidate set for downlink training phase (Number of transmit antennas $N_t=16$, number of RIS elements $N=64$, SNR = 10dB, proposed algorithm threshold $\gamma = 0.9$, and walking speed $v=0.6m/s$).}
\label{fig: fig5}\vspace{-0.0 cm}
\end{figure}

In order to reflect the superior of our proposed algorithm in saving signaling time, in Fig. \ref{fig: fig5}, we examine the effect of the size of candidate set for the optimization problem (\ref{eq: obj two errors}). 
We evaluate the average error rate for the proposed algorithm, calculating by the difference of instantaneous rate between the oracle idea, i.e., $|R_{ins}^{(t)} - R_{Orc}^{(t)}|$.
It can be seen that the error of average rate converges very quickly, with near-zero error when the size is larger than 5, which verifies that our proposed algorithm can dramatically save signaling time, comparing with conventional exhaustive search idea.
However, in order to make sure the optimal solution is always obtained inside the found candidate set, we set the size of the candidate set as 7 in all other simulation studies.

\subsection{Complexity Comparison}
\label{sec:complexity}

\begin{table}
\caption{Percentage of time-slots below corresponding thresholds.}
\begin{center}
 \begin{tabular}{ccccc}
\toprule
 ~ & Prop. & Exh, res = $1^\circ$ & Exh, res = $5^\circ$ &  Exh, res = $10^\circ$\\
\midrule
 1 & $0.32\%$ & $4.53\%$ & $0.54\%$ & $0.21\%$ \\ 
 2 & $0.5\%$ & $7.47\%$ & $0.94\%$ & $0.39\%$ \\
 3 & $0.8\%$ & $11.89\%$ & $1.54\%$ & $0.55\%$ \\
\bottomrule
\end{tabular}
\label{tab: timeslots}
\end{center}
\end{table}

Finally, we turn to evaluate the complexity of our proposed beam tracking algorithm (Prop.) comparing with the conventional exhaustive searching idea (Exh.) with different resolutions.
In Tab. \ref{tab: timeslots}, we summarized the percentage of time-slots below corresponding thresholds for both algorithms.
We implement this table with 3 channel realizations.
The data in the first row is generated under the channel condition that the distances of AP-RIS and RIS-UE are both $r_1 = r_2 = 2m$ with user velocity as $v=0.6m/s$.
The second row is when the distances are set as both $r_1 = r_2 =3m$ with user velocity as $v=1.2m/s$, and the third row is when the distances are set as both $r_1 = r_2 =4m$ with user velocity as $v=1.8m/s$.
We calculate the number of time-slots below corresponding thresholds under three cases:
1) Number of time-slots in DL training phase when updating RIS configuration: 7 in the proposed algorithm, 360 in exhaustive searching idea with res = $1^\circ$, 72 in exh, res = $5^\circ$, and 36 in exh, res = $10^\circ$;
2) Number of time-slots when DL data transmission is below the thresholds: 1 for both ideas;
3) Number of time-slots when conducting UL feedback: 1 for indicating the start of the beam tracking algorithm and 1 for indicating its end for both ideas.

From the table, we can see that the proposed algorithm has a percentage of time-slots below the threshold $\gamma=0.9$ as small as $0.32\%$ while this percentage for exhaustive searching idea with resolution $1^\circ$ is $4.53\%$.
The accuracy of the RIS configuration for exhaustive searching idea is increasing with finer resolution at the cost of increasing the signaling time in DL training as is seen when we comparing the same row of exhaustive searching idea with different resolution.
Therefore, we can conclude that, in case of comparable instantaneous rate, our proposed algorithm has over 14 times signaling time saved than the conventional exhaustive searching idea.

In addition, when we compare different rows, it is verified that with higher user velocity, the percentage of signaling time used for beam tracking will get higher.
Such observation also verifies the conclusion drawn for Fig. \ref{fig: fig4}.

\begin{table}
\caption{Number of beam tracking procedure called}
\begin{center}
 \begin{tabular}{ccccc}
\toprule
 ~ & Prop. & Exh, res = $1^\circ$ & Exh, res = $5^\circ$ &  Exh, res = $10^\circ$\\
\midrule
 1 & 101 & 40 & 23 & 17 \\ 
 2 & 80 & 33 & 20 & 16 \\
 3 & 86 & 35 & 22 & 15 \\
\bottomrule
\end{tabular}
\label{tab: called}
\end{center}
\end{table}

We also illustrate the number of beam tracking procedure called for both ideas in Tab. \ref{tab: called}.
As is shown in the table, a more frequent calling of the beam tracking procedure results in a higher achievable rate.
This is because that the effectiveness of RIS configuration will be deteriorated by the mobility of the user and a more frequent calling of the beam tracking procedure will adjust the RIS configuration more frequently, thus maintains the instantaneous rate at a rather high level.

\section{Conclusion}
\label{sec: conclusion}

In this paper, we propose a fast RIS-based beam tracking algorithm in a mmWave system.
We first exploit the differential form of optimal RIS configuration and serve it as the updating beam tracking parameter to avoid complex channel estimation procedure.
Then we transform the RIS-based beam tracking problem into an optimization problem solved by a two-dimensional grid-based search.
Finally, with a small set candidate solutions, the downlink signaling time can be dramatically saved.
Simulation studies showed that the proposed algorithm outperforms exhaustive searching idea and has near-optimal performance.

\end{document}